\documentclass[12pt]{article}
\pdfoutput=1

\usepackage{amssymb,amsmath,slashed}
\usepackage[pdftex]{graphicx}
\makeatletter

\@addtoreset{equation}{section}
\def\section{\@startsection {section}{1}{\z@}{-2.5ex plus -1ex minus
 -.2ex}{1.3ex plus .2ex}{\large\bf}}
\def\subsection{\@startsection{subsection}{2}{\z@}{-2.25ex plus%
 -1ex minus -.2ex}{0.5ex plus .2ex}{\bf}}

\advance \voffset by -0.9in
\advance \hoffset by -0.75in
\newcommand{\MM}{\mathcal{M}}

\newcommand{\CP}{\mathbb{CP}}

\newcommand{\RR}{\mathbb{R}}
\newcommand{\ZZ}{\mathbb{Z}}

\newcommand{\bee}{\begin{equation}}
\newcommand{\eee}{\end{equation}}

\begin{document}

\begin{flushright}
EMPG-14-22
\end{flushright}
\vskip 10pt
\baselineskip 28pt

\begin{center}
{\Large \bf  Taub-NUT Dynamics with a  Magnetic Field}

\baselineskip 18pt

\vspace{1 cm}

{\bf Rogelio Jante and Bernd J.~Schroers},\\
\vspace{0.2 cm}
Maxwell Institute for Mathematical Sciences and
Department of Mathematics,
\\Heriot-Watt University,
Edinburgh EH14 4AS, UK. \\
{\tt rj89@hw.ac.uk} and {\tt b.j.schroers@hw.ac.uk} \\
\vspace{0.4cm}

{ July  2015} 
\end{center}

\begin{abstract}
\noindent  We study classical and quantum dynamics on the Euclidean Taub-NUT geometry coupled to an abelian gauge field with self-dual curvature and show that, even though Taub-NUT has  neither bounded orbits nor  quantum bound states, the magnetic binding via  the gauge field produces both. The conserved Runge-Lenz vector of Taub-NUT dynamics survives, in a modified form,  in the gauged model and allows for an essentially algebraic computation of classical trajectories and  energies of  quantum bound states.  We also compute scattering cross sections  and find  a surprising electric-magnetic duality. Finally, we exhibit the dynamical  symmetry behind the conserved Runge-Lenz and angular momentum vectors in terms of a twistorial formulation  of phase space.  
\end{abstract}

\baselineskip 16pt
\parskip 5 pt
\parindent 0pt

%
%

\section{Introduction}

The Euclidean Taub-NUT (TN) geometry has been studied extensively and from several different points of view. It is interesting  as a particularly simple example of a gravitational instanton,  it can be viewed as a Kaluza-Klein geometrisation of the Dirac monopole and it arises in the context of monopole moduli spaces, either directly, or, in its `negative mass' form, as an asymptotic limit. In each of  these contexts it plays a role akin to   that  of the hydrogen atom in atomic physics, both in  the general sense of being the simplest example and in the  technical sense of sharing simplifying features with the hydrogen atom, like   a Runge-Lenz type conserved quantity.

The four-dimensional Maxwell equations on TN space have a simple  source-free solution which is intimately connected to the TN geometry.  This was first pointed out by   Pope \cite{Pope1,Pope2} who went on to show that that the index of the Dirac operator minimally coupled to this Maxwell field is non-trivial. The index and the properties of  zero-modes of this gauged Dirac operator were recently studied in detail in our  paper \cite{JS} from which the current paper evolved. 

The Maxwell field  first considered by Pope has   played  an important role in various contexts.  
Its field strength is a harmonic two-form  which is self-dual for an appropriate choice of orientation and also square-integrable. It is exact,  with a globally defined gauge potential which is, however, not square-integrable. In other words, the harmonic two-form generates the  non-trivial $L^2$-cohomology in the middle dimension of TN space. This is the reason why it was important   in tests of S-duality on monopole moduli spaces \cite{Weinberg,Gauntlett}.  

One can relate the self-dual two-form  directly to the TN geometry by noting that, with a suitable normalisation, it is  the Poincar\'e dual of the $\CP^1$ which compactifies TN to $\CP^2$  \cite{AMS}. More generally, one can understand  the $L^2$-cohomology of TN and  its multi-centre generalisation in terms of   the ordinary cohomology of a suitable compactification   \cite{Guido}.

 For all of these reasons, it is not surprising that the inclusion of the  self-dual gauge field in the dynamics on TN space turns out to be mathematically natural.  In this paper, we consider the geodesic motion on  TN coupled to the self-dual gauge field  as our model for the classical dynamics and   the Dirac and Laplace operators on TN, also  minimally coupled to the gauge field, as quantum models. We   show that all the  interesting  algebraic features of ordinary TN dynamics carry over to the gauged case, and that, moreover, bounded motions and quantum bound  states, neither of which are possible on (regular, `positive mass') TN alone,  occur in the gauged dynamics.

In general qualitative  and semi-quantitative terms, the classical and quantum dynamics on TN space is strikingly similar to the Kepler problem and the non-relativistic hydrogen atom. 
The basic reason why the   four-dimensional TN geometry can model the three-dimensional motion of a charged particle is the  fact that TN is a Kaluza-Klein geometrisation of the  Dirac monopole in three dimensions \cite{Sorkin,GP}.  The more detailed similarities are related  to  the conserved Runge-Lenz vector in both cases. In this sense, one can view our inclusion of the magnetic field in the TN dynamics as analogous to the inclusion of a magnetic monopole field in the  much studied extension of the Kepler problem to the MICZ Kepler problem \cite{MC,Zwanziger}.
 
From the Kaluza-Klein point of view, the magnetic field of the monopole is already encoded in the geometry, and the name  `magnetic field' for an additional abelian gauge field on TN is potentially confusing. We adopt it here because it is justified  from the four-dimensional point of view. As we shall see,  the magnetic  field leads to magnetic binding  akin to that responsible for Landau levels in planar systems. In fact, in a limit where TN becomes flat Euclidean four-space, the four-dimensional magnetic  field is constant, and the bound states that we find become ordinary Landau levels.  This picture of magnetic binding also provides a qualitative explanation of the index found by Pope and of  the form of the zero-modes discussed in \cite{JS}.

 We have organised our presentation to proceed from the most direct to more abstract treatments of gauged  TN dynamics. We begin in Sect.~2  with a  brief general discussion  of how a magnetic field on a two-dimensional  Riemannian manifold can produce bound states. In Sect.~3, we collect conventions for describing the  TN  geometry and the associated Dirac and Laplace operators coupled to a magnetic field.   We   turn to the classical dynamics  in Sect.~4, discuss  the conserved Runge-Lenz and angular momentum vectors   of the gauged geodesic motion on TN and describe the classical trajectories. Sect.~5 contains a direct solution of the eigenvalue problem for  the  gauged Laplace operator  on  TN space  through separation of variables. We    exhibit the promised bound  states, give their energies and  degeneracies and  compute  scattering cross sections.   In Sect.~6, we solve the quantum problem algebraically, using a quantum version of the Runge-Lenz vector. In Sect.~7, we exhibit the symmetry underlying the conservation of angular momentum and the Runge-Lenz vectors from a twistorial description of phase space. Our final Sect.~8 contains a brief discussion, our conclusions and an outlook onto open problems.
 
\section{A toy model: motion on a  surface  with magnetic field}
\label{toysect}

We can  gain a qualitative understanding of   bound states  on TN coupled to a Maxwell field   by considering  a two-dimensional model, consisting of a two-dimensional manifold with metric and magnetic field. We will encounter a manifold and metric of the same kind  in our study of TN as a geodesic submanifold, and the magnetic field as the restriction of the Maxwell field to the geodesic submanifold. However, here we study  the two-dimensional model in its own right. 

Consider a two-dimensional manifold diffeomorphic to an open disk $D$  with $U(1)$-invariant metric of the form 
\bee
\label{toy}
ds^2 =dR^2 + c^2(R) d\gamma^2.
\eee
For consistency with our later discussion of the TN geometry we take the angular coordinate $\gamma$ in the interval $[0,4\pi)$, so that $4\pi c$ is the length of a  $U(1)$ orbit. The radial coordinate $R$ is the proper  radial distance from the origin and has range $[0,\infty)$, and we assume a form of  $c$ near $R=0$ to ensure that the metric is smooth there.   We are interested in two kinds of behaviour of the function $c$.

\begin{figure}[!h]
\centering
\includegraphics[width=8truecm]{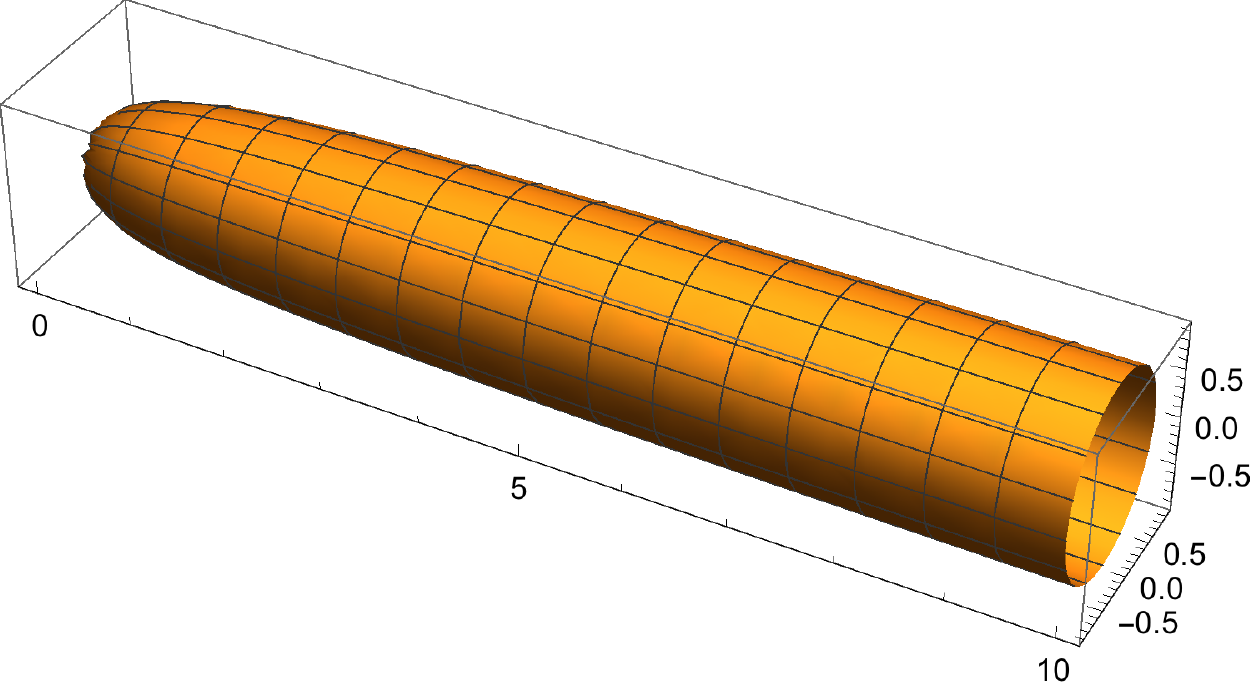} \quad \quad  \quad \quad 
\includegraphics[width=4truecm]{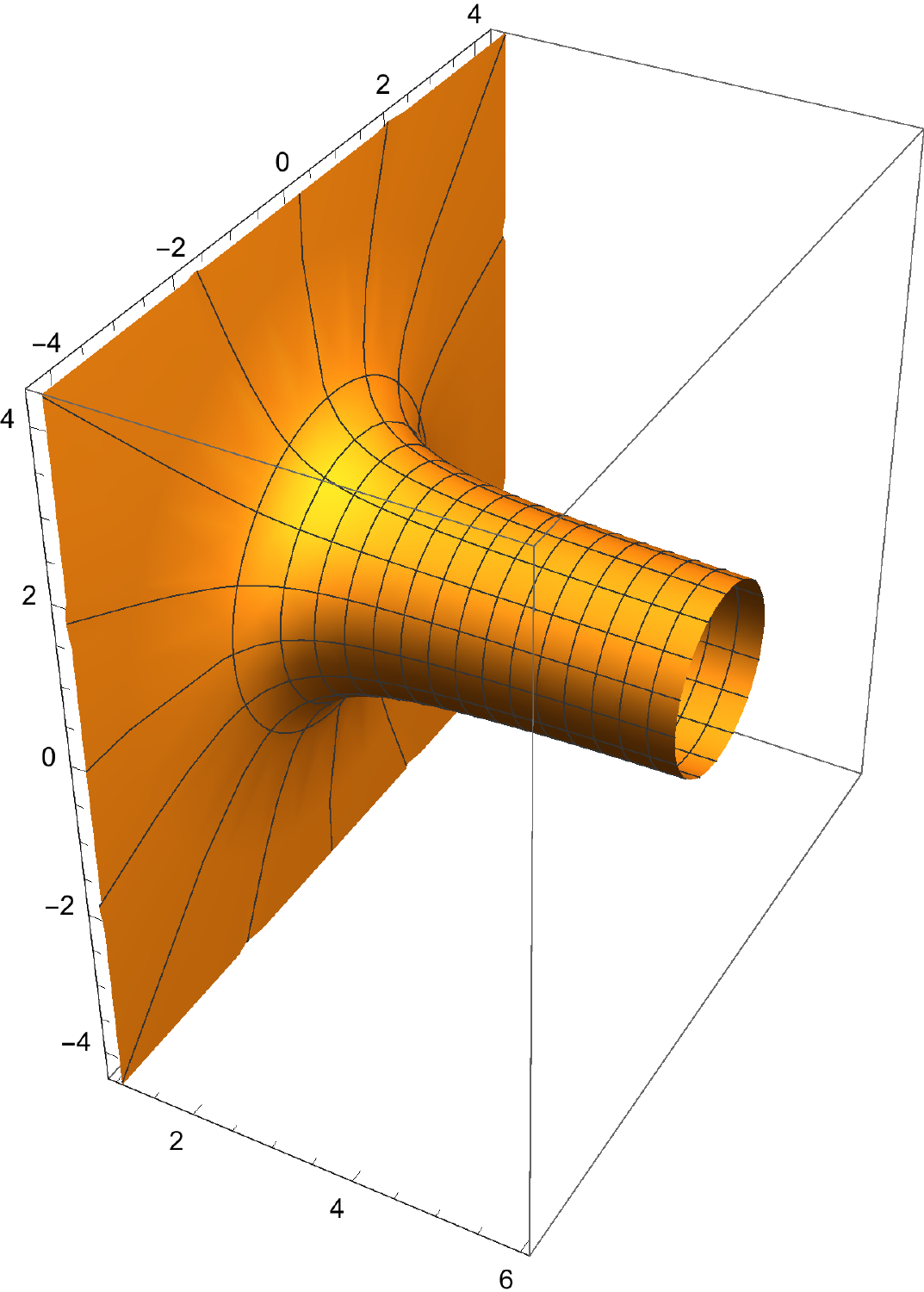}
\caption{The cigar-shaped surface  for  positive $L$ (left) and the funnel-shaped surface for  negative $L$ (right).  }
\label{cigarfunnel}
\end{figure}

The first case captures what happens in the regular TN geometry. The function $c$ has the finite range $[0,L)$  for some positive real number $L$ so that the length of the $U(1)$ orbits remain bounded.  Moreover we assume that $c(0)=0$ and that $c$ is strictly monotonic, so that  one can picture the metric as being induced on a  cigar-shaped surface of revolution in three-dimensional Euclidean space, as shown in Fig.~\ref{cigarfunnel}. The qualitative behaviour of geodesics on such a surface is well know and follows from Clairaut's relation. Generic geodesics spiral on the cigar. Geodesics spiralling towards the tip will be reflected at some point and spiral out.   All geodesics ultimately move arbitrarily far away from the tip and there are no geodesics which remain in a region  bounded by a finite value of $R$.

The second case captures what happens in  the singular or `negative mass'  TN. The function $c$ diverges at $R=0$, has the range $(L,\infty)$  and is monotonically decreasing. As an embedded surface, this is a funnel, with the opening at $R=0$ and the tip at $R=\infty$ as shown in Fig.~\ref{cigarfunnel}. Generic geodesics again spiral on this surface, but now there are two kinds of behaviour. Geodesics which travel straight down the funnel or spiral only slowly may escape to  $R=\infty$. However, geodesics  travelling into the funnel with sufficiently high angular momentum relative to their speed will bounce back and remain inside a region bounded by some finite value of $R$. 

We now return to the first case with monotonically increasing $c\in [0,L)$ and  consider the inclusion of a magnetic field of a specific type given  by the two-form
\bee
\label{magfield}
B= d\left( \frac{pc^2}{2L^2} d\gamma\right) = \frac{p}{L^2} c\, dc \wedge d\gamma,
\eee
for some real constant $p$ which controls the strength of the magnetic field, and is proportional to its flux:
\bee
\frac {1}{2\pi} \int_D B = p. 
\eee
The Lagrangian governing the motion of particle on the surface with  metric \eqref{toy}, minimally coupled to the gauge potential for  $B$ is, for a suitably chosen mass parameter, 
\bee
\mathcal{L} = \frac 1 4 \left ( \dot{R}^2 +c^2 \dot{\gamma}^2 \right) - \frac{pc^2}{2L^2} \dot{\gamma}.
\eee
With the momenta conjugate to $R$ and $\gamma$ 
\bee
p_R=\frac{\partial \mathcal{L}}{\partial \dot{R}} =\frac{1}{2}\dot{R} \qquad q=\frac{\partial \mathcal{L}}{\partial \dot{\gamma}}=\frac 1 2  c^2\dot{\gamma} - \frac{pc^2}{2L^2}, 
\eee
the Hamiltonian is 
\bee
H =  p_R^2 + \left(\frac{q}{c}+\frac{pc}{2L^2}\right)^2.
\eee
Since $q$ is conserved and $p$ constant, this is effectively the Hamiltonian for one-dimensional motion on the half-line  in the potential
\bee
\label{simplepot}
W=\left(\frac{q}{c}+\frac{pc}{2L^2}\right)^2. 
\eee

\begin{figure}[!h]
\centering
\includegraphics[width=8.5truecm]{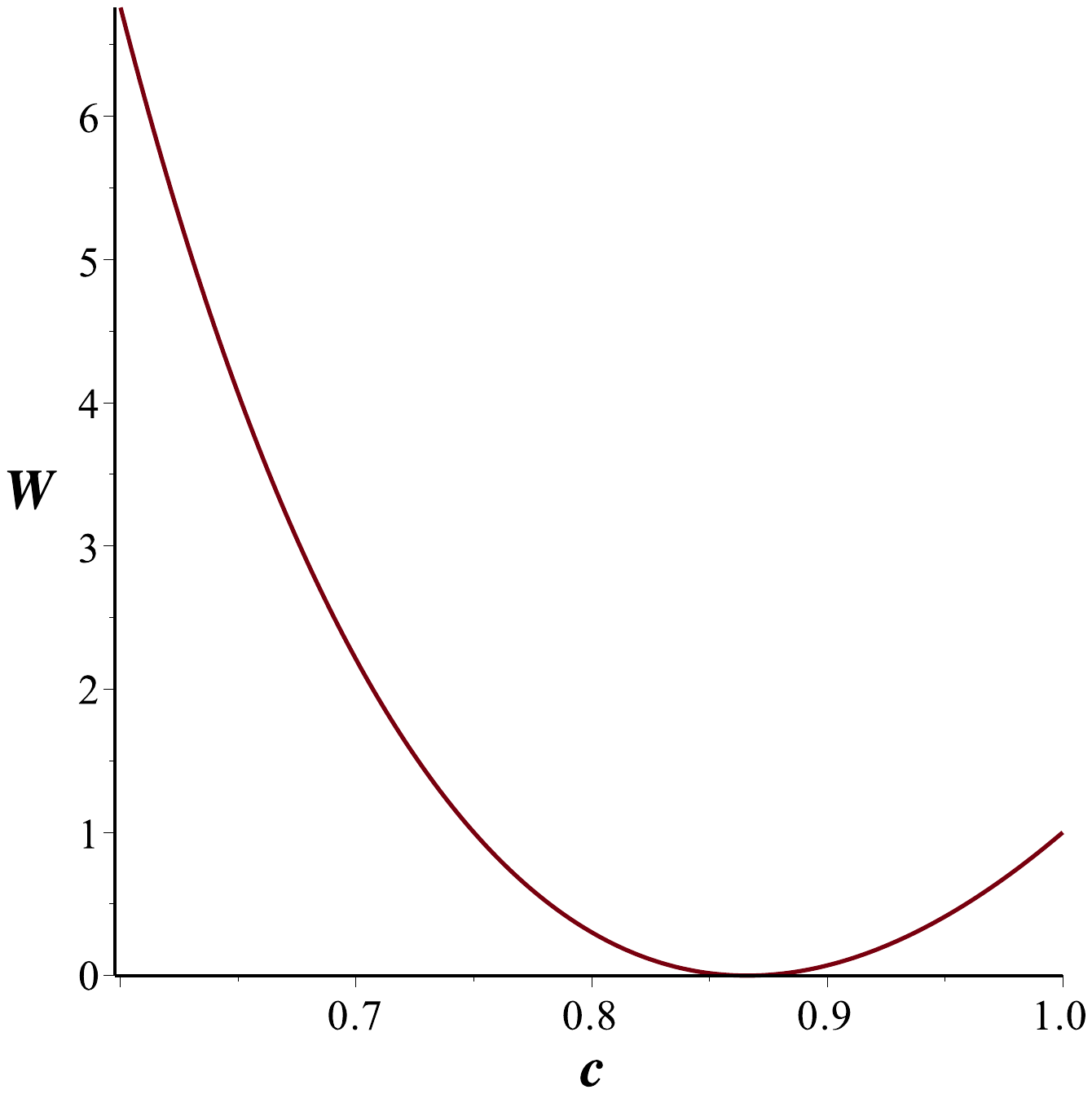}\hspace{-1cm}
\includegraphics[width=8.5truecm]{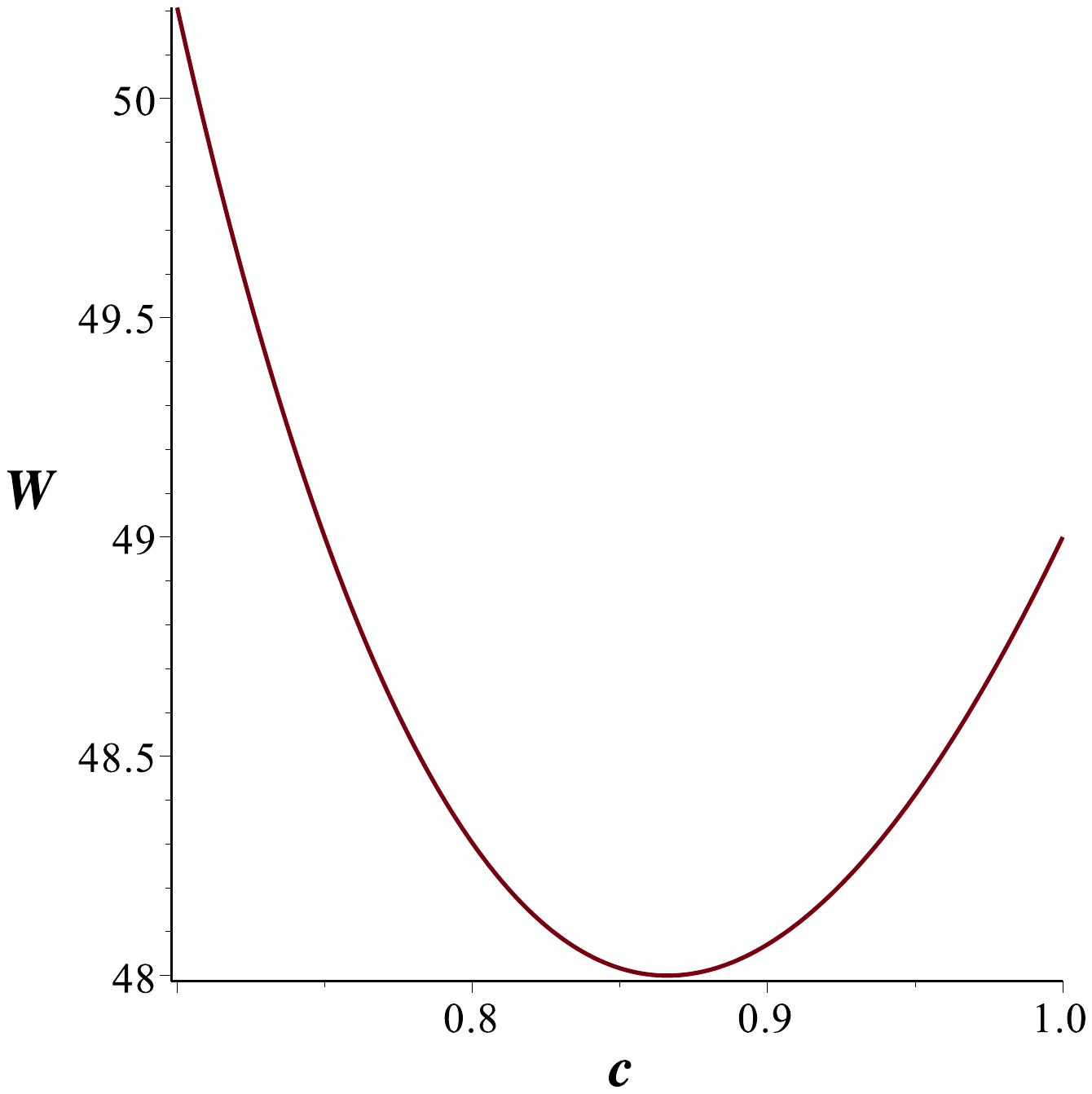}
\vspace{-5cm}
\caption{Plots of the potential \eqref{simplepot} for $L=1$ and $c\in [0,1)$ for $q=3$ and $p=-8$ (left) and $q=3$ and $p=8$ (right).}
\label{simpleplots}
\end{figure}

We would like to know if there are bounded trajectories in the potential \eqref{simplepot}.  As a potential, $W$ should be viewed as a function of $R$, but with our assumption that $dc/dR > 0$ we can study its minima by looking at $W$ as a function of $c$.  It is easy to check that $W(c)$ has a unique minimum  at  $c_m>0$ satisfying
\bee
\frac{c_m^2}{L^2} = \left|\frac{2q}{p}\right|. 
\eee
However, for  $c_m$ to be in the range $[0,L)$ we require
\bee
\label{toycond}
|q|< \left|\frac p 2\right|,
\eee
and this is a necessary and sufficient condition for $W$ to have a minimum.
The value at the minimum is 
\bee
\label{toymin}
W(c_m)=\begin{cases}  0 & \text{if} \; pq < 0 \\
\frac{2pq}{L^2} & \text{if} \; pq \geq 0. \end{cases}
\eee
The qualitative form of the potential is similar in the two cases. We also note that the inequality   \eqref{toymin} implies the bound
\bee
\label{toybound}
L^2 W \geq \begin{cases}  0 & \text{if} \; pq < 0 \\
2pq & \text{if} \; pq \geq 0, \end{cases}
\eee
which will play an important role in our discussion. 

 We conclude that a magnetic field on a cigar-shaped surface can lead to bounded trajectories even though the geometry of the cigar does not support any bounded geodesics.  In the presence of  the magnetic field \eqref{magfield}, all  trajectories with  angular momentum $q$ of magnitude less than $|p/2|$ remain in a bounded region.

 \section{Classical and quantum geometry of the Taub-NUT space}
 \subsection{The Taub-NUT geometry}
 \label{TNreview}
We use the conventions of  \cite{JS}  for the description of the TN geometry, and refer the reader to that paper for details. Here we summarise essential background and notation. 
The TN space $M_{\text{\tiny TN}}$ is a non-compact four-dimensional Riemannian manifold with a self-dual Riemann tensor and $U(2)$ isometry.  As a manifold it is diffeomorphic to $\RR^4$,  and  the   metric can be written as
\begin{equation}\label{TN}
ds^2 = f^2dr^2 + a^2\sigma_1^2 + b^2\sigma_2^2 + c^2\sigma_3^2,
\end{equation} 
where  $\sigma_j$ are the left-invariant one-forms on $ SU(2)\simeq S^3$. They are defined   in terms of $h\in SU(2)$  via
\begin{equation}\label{sig}
h^{-1}dh = \sigma_1t_1 + \sigma_2t_2 + \sigma_3t_3, 
\end{equation}
where we use $su(2)$ generators which are given in terms of the Pauli matrices $\tau_j$ as  $t_j = -\frac{i}{2}\tau_j$.

The quantities $a,b,c, f$ are functions of a  radial  coordinate $r$ transverse to the $SU(2)$ orbits. The self-duality of the  Riemann curvature is then equivalent to 
\begin{equation}
\frac{2bc}{f}\frac{da}{dr} = (b-c)^2 - a^2, + {\rm \ cycl.}\;,
\end{equation}
and the  solution which  gives rise to the TN metric is 
\begin{equation}\label{solu}
a = b = r\sqrt{V}, \ \ c = \frac{L}{\sqrt{V}}, \ \  f = -\frac{b}{r}, \ \ V = \epsilon + \frac{L}{r}.
\end{equation}

Here $\epsilon$ and $L$ are parameters which are required to be positive for a smooth metric. As discussed in \cite{AFS}, the parameter $\epsilon$ is relevant  for taking the Landau limit of our model, and can be used to introduce time-dependence in an interesting way. However, we shall set $\epsilon=1$ for most of the text and work with the  potential 
\bee
V =  1 + \frac{L}{r},
\eee
unless stated otherwise. The TN space with $L<0$  arises as the asymptotic form of the two monopole moduli space \cite{GM} but  has a singularity at $r = -L$. We will assume $L>0$ in the following.

For the discussion of symmetries and the definition of natural differential operators associated to the TN geometry,  we require the  vector fields  $X_j$ which are  dual to  the left-invariant forms $\sigma_j$ (so $\sigma_j(X_i)=\delta_{ij}$) and which generate  right-actions on $h\in SU(2)$,
\begin{equation}\label{h}
X_j: h \to ht_j. 
\end{equation}
We will also need the vector fields  $Z_j$ which generate  the  left-action,
\begin{equation}\label{LA}
Z_j: h \to -t_jh.
\end{equation}
It follows from these definitions that, as vector fields on $S^3$,  
\bee
[X_i,X_j]=\epsilon_{ijk} X_k, \qquad  [Z_i,Z_j]=\epsilon_{ijk} Z_k, \qquad [X_i,Z_j]=0,
\eee
where summation over repeated indices is assumed.  By definition, the Lie derivative of the  one-forms $\sigma_i$ with respect to the $Z_j$ vanishes.

There are several natural  ways to parametrise  $SU(2)$ and hence to write the one-forms and vector fields on it. For the purposes of this paper we need both a parametrisation in terms of Euler angles and one in terms of two complex numbers. Continuing with the conventions of \cite{JS} we therefore introduce complex numbers $z_1,z_2$, constrained to satisfy $|z_1|^2+|z_2|^2=1$, and write elements of $SU(2)$ as 
\bee
h=\begin{pmatrix} z_1 & -\bar{z}_2 \\ z_2 & \phantom{-} \bar{z}_1 \end{pmatrix}.
\eee 
In some calculation it is more convenient to work with Euler angles  $\beta \in [0,\pi], \alpha \in [0,2\pi), \gamma \in [0,4\pi)$  defined via
\begin{equation}
\label{s3}
z_1= e^{-\frac{i}{2}(\alpha + \gamma)}\cos{\frac{\beta}{2}}, \ \ z_2 = e^{\frac{i}{2}(\alpha - \gamma)}\sin{\frac{\beta}{2}}.
\end{equation}

Then the left-invariant one-forms can be expressed in terms of Euler angles as 
\begin{eqnarray}\label{LI}
 \sigma_1 &=&  \sin\gamma d\beta - \cos\gamma \sin\beta d\alpha,  \nonumber\\
 \sigma_2 &=& \cos\gamma d\beta + \sin\gamma \sin\beta d\alpha, \nonumber\\
\sigma_3 &=&  d\gamma  + \cos\beta d\alpha ,
\end{eqnarray}
and the dual vector fields are 
\begin{align}\label{dual}
 X_1 &= \phantom{-}\cot\beta \cos \gamma \partial_\gamma
           + \sin \gamma \partial_\beta
            -\frac {\cos \gamma }{ \sin\beta} \partial_\alpha,
\nonumber\\
X_2 &= -\cot\beta \sin \gamma \partial_\gamma
           + \cos \gamma \partial_\beta
            + \frac {\sin \gamma}{ \sin\beta} \partial_\alpha,
\nonumber \\
 X_3 &= \phantom{-}\partial_\gamma.
\end{align}

In terms of complex coordinates we have 
\begin{align}
\label{bodyfixed}
X_1 &= \frac{i}{2}(\bar{z}_2\partial_1 -\bar{z}_1\partial_2 - z_2\bar{\partial}_1 + z_1\bar{\partial}_2), \nonumber \\
X_2 &= \frac{1}{2}(-\bar{z}_2\partial_1 + \bar{z}_1\partial_2 - z_2\bar{\partial}_1 + z_1\bar{\partial}_2), \nonumber \\
X_3 &= \frac{i}{2}(-z_1\partial_1 -z_2\partial_2 + \bar{z}_1\bar{\partial}_1 + \bar{z}_2\bar{\partial}_2),
\end{align}
and 
\begin{align}
\label{spacefixed}
Z_1 &= \frac{i}{2}(-z_2\partial_1 -z_1\partial_2 + \bar{z}_2\bar{\partial}_1 + \bar{z}_1\bar{\partial}_2), \nonumber \\
Z_2 &= \frac{1}{2}(-z_2\partial_1 + z_1\partial_2 - \bar{z}_2\bar{\partial}_1 + \bar{z}_1\bar{\partial}_2), \nonumber \\
Z_3 &= \frac{i}{2}(z_1\partial_1 -z_2\partial_2 - \bar{z}_1\bar{\partial}_1 + \bar{z}_2\bar{\partial}_2). 
\end{align}
It follows from general arguments,  but can also be checked explicitly,  that the Laplace operator on $S^3\simeq SU(2)$  can be written as 
\bee
\label{LaplaceSU2}
\Delta_{S^3} = Z_1^2+Z_2^2+Z_3^2=  X_1^2+X_2^2+X_3^2.
\eee

The magnetic field on TN  referred to in the title and discussed in the Introduction can be written as the  exterior derivative of the  abelian  gauge field
\begin{equation}
\label{A}
 \mathcal{A} = \frac{pc^2}{2L^2}\sigma_3 = \frac{p}{2}\frac{r}{r+L} \sigma_3,
\end{equation}
where $p$ is a real parameter which will play an important role in our discussions.
The magnetic field  
\bee
\label{FA}
\mathcal{F} =d \mathcal{A} =\frac{p}{L^2}\left(c dc\wedge \sigma_3 + \frac{c^2}{2}\sigma_2\wedge \sigma_1\right),
\eee
 is (up to a multiplicative constant)  the unique harmonic, normalisable and $U(2)$-invariant two-form on TN.

Note that the gauge we have chosen preserves the $U(2)$ symmetry of the TN geometry. However, the magnetic field  inevitably breaks the discrete symmetry 
\bee
\label{broken}
 \beta \mapsto \pi -\beta, \quad \alpha\mapsto \pi + \alpha, \quad \gamma \mapsto -\gamma,
\eee
which maps 
\bee
\sigma_1\mapsto \sigma_1, \quad \sigma_2\mapsto -\sigma_2, \quad \sigma_3 \mapsto -\sigma_3,
\eee
and therefore preserves the metric \eqref{TN} but neither  the gauge field \eqref{A} nor its curvature. As we will see,  this has  interesting consequences in the dynamics. 

 TN without the point $r=0$ (the NUT)  is a circle bundle over $\RR^3\setminus\{0\}$.  As reviewed in \cite{AFS}, for each direction $(\beta,\alpha)$ in the base there is a  geodesic submanifold parametrised by $(r,\gamma)$. Each of these geodesic submanifolds is of the general cigar-shape of our toy model in Sect.~2, and the flux of the magnetic field  \eqref{FA} through this submanifold is $2\pi p$.  Thinking of TN as a two-sphere's worth of such cigar-shaped surfaces threaded by magnetic flux will prove very helpful for a qualitative understanding of our results.

\subsection{The gauged Dirac and Laplace operators}

The most fundamental operator associated to the metric \eqref{TN} and the connection \eqref{A} is the Dirac operator on TN minimally coupled to the connection.  As shown in  \cite{CH} for the ungauged case, the spectrum of the Dirac operator and the Laplace operator are closely related. The arguments in \cite{CH} are essentially a reflection of an underlying supersymmetry. It is not difficult to adapt them to the gauged case, as we shall now show. 
 
The  gauged Dirac operator  on TN has the  following form \cite{JS}
\begin{equation}
 \slashed{D}_p = \left(\begin{array}{cc}
 0 & T_p^{\dagger} \\
 T_p & 0
 \end{array}\right),
\end{equation}
where
\begin{align}
T^\dagger_p&= \frac{i}{\sqrt{V}}\left(-\partial_r  -\frac 1 r -   \frac{V}{2L} +\tau_3 \left(\frac{p}{2L} - \frac{iV}{L} X_3\right) -\frac i r (\tau_1X_1 +\tau_2X_2)\right), \nonumber \\
T_p & = \frac{i}{\sqrt{V}}\left(-\partial_r  -\frac 1 r +\frac{V}{2L} + \frac{L}{2r^2V}
+\tau_3 \left(\frac{iV}{L}X_3-\frac{p}{2L}\right)+ \frac i r (\tau_1 X_1 +\tau_2X_2)\right). 
\end{align}

 As shown in \cite{Pope1,Pope2}  and elaborated in \cite{JS}, the kernel of $T^\dagger_p$ is trivial but that of $T_p$ has dimension
 $\frac 1 2 [|p|] ([|p|]+1)$,
where $[x]$ is the largest integer {\em strictly} smaller than  the positive real number $x$.  It follows that 
 \bee
 H_- = T_pT_p^{\dagger}
\eee
is a strictly positive operator, and that 
 \bee
 H_+ = T_p^{\dagger}T_p
 \eee
 is positive, but not strictly positive. Therefore,  we can define the unitary operator
\bee
U = \frac{1}{\sqrt{H_-}}T_p,
\eee
with inverse $ U^{-1} = T_p^{\dagger}/\sqrt{H_-}$,  and use it to relate $H_+$ and $H_-$ via 
\bee
 H_+ = U^{-1}H_-U.
\eee
It follows from this unitary equivalence that, as in the ungauged case \cite{CH}, $H_+$ has the same spectrum as $H_-$ apart from the zero eigenvalue of $H_-$. In other words if $\Psi$ is a two-component eigenspinor of $H_-$ with eigenvalue $E$ then $U^{-1}\Psi$ is an eigenspinor of $H_+$ with the same eigenvalue. 

Combining these results, we  obtain eigenstates of the Dirac operator  with non-zero eigenvalues from eigenstates  of $H_-$ as follows. Suppose that  $\Psi$ is an  eigenstate  of $H_-$ with eigenvalue $E>0$. Then 
 \begin{equation}
 \slashed{D}\binom{U^{-1}\Psi}{\pm\Psi} = \pm\sqrt{E}\binom{U^{-1}\Psi}{\pm\Psi}.
 \end{equation}

Inserting the expressions for the TN profile functions, we find
 \bee
H_- = T_{p}T^{\dagger}_{p} =-\frac{1}{r^2V}\partial_r(r^2\partial_r)  - \frac{1}{r^2V}\left[(X_1 + t_1)^2+ (X_2 + t_2)^2\right] -\frac{V}{L^2}\left(X_3 + \frac{ip}{2V}+ t_3\right)^2.
\eee
It is convenient to change gauge by observing  from (\ref{h}) that
 \begin{equation}
 X_ih^{-1} = -t_ih^{-1}
 \end{equation}
and that therefore, as an operator identity, 
 \begin{equation}
 h(X_i + t_i)h^{-1} = h(X_ih^{-1}) + hh^{-1}X_i + ht_ih^{-1} = X_i.
 \end{equation}
Employing this,  we obtain
\begin{equation}
hH_-h^{-1} = H_p\tau_0,
\end{equation}
where $\tau_0$ is the $2\times 2$ identity matrix and
\begin{equation}\label{Hp}
H_p = -\frac{1}{r^2V}\partial_r(r^2\partial_r) - \frac{1}{r^2V}(X_1^2 + X_2^2) - \frac{V}{L^2}\left(X_3 + \frac{ip}{2V}\right)^2
\end{equation}
is the Laplace operator associated to the metric \eqref{TN} and minimally coupled to the gauge field \eqref{A}. This is the operator whose spectrum we shall study in the remainder of this paper. 
 
\section{Dynamical symmetries in classical  Taub-NUT dynamics}
\subsection{Canonical procedure}
We now turn our attention to the classical dynamics in TN in the gauged case. We discuss the conserved angular momentum  and  Runge-Lenz vectors, and use them to describe the classical trajectories.  Our treatment is an extension of the discussion  in  \cite{GM} and \cite{FH} of the (ungauged) motion on TN space.

As reviewed in Sect.~2,  the TN space (\ref{TN}) can be parametrised by a coordinate $r$ and the Euler angles $ \alpha, \beta$ and $\gamma$. 
In these coordinates the Lagrangian  for geodesic motion on TN takes the form 
\begin{equation}
\mathcal{L} = \frac{1}{4}(f^2\dot{r}^2 + a^2\omega_1^2 + b^2\omega_2^2 + c^2\omega_3^2),
\end{equation}
where $\omega_i$ are the components of the body fixed angular velocity,
\begin{eqnarray}
\omega_1 &=& \sin{\gamma} \,\dot{\beta} - \cos{\gamma}\sin{\beta}\,\dot{\alpha}, \nonumber \\
\omega_2 &=& \cos{\gamma}\, \dot{\beta} + \sin{\gamma}\sin{\beta}\,\dot{\alpha}, 
\nonumber \\
\omega_3 &=& \dot{\gamma} + \cos{\beta}\,\dot{\alpha},
\end{eqnarray}
and we have chosen an overall factor of $1/4$ for convenience.

Inserting the angular velocities in the Lagrangian and recalling that $a = b = r\sqrt{V}, c = L/\sqrt{V}, f = -b/r$ we obtain
\begin{equation}
\mathcal{L} = \frac{1}{4}\left[V(\dot{r}^2 + r^2\dot{\beta}^2+r^2\sin^2{\beta}\dot{\alpha}^2) + L^2V^{-1}(\dot{\gamma} + \cos{\beta}\dot{\alpha})^2\right].
\end{equation}
In terms of cartesian coordinates
\bee
\vec{r} = (x_1,x_2,x_3) = (r\sin \beta \cos \alpha, r \sin\beta \sin\alpha , r\cos \beta),
\eee 
the Lagrangian takes the more familiar form 
\begin{equation}
\mathcal{L} = \frac{1}{4}(V |\dot{\vec{r}}|^2 + L^2V^{-1}(\dot{\gamma} + \vec{A}\cdot\dot{\vec{r}})^2),
\end{equation}
where  $\vec{A}$ is a gauge potential for the Dirac monopole
 \begin{equation}\label{Dmonopole}
A_1 = -\frac{x_3x_2}{r(r^2-x_3^2)}, \ \ A_2 = \frac{x_3x_1}{r(r^2-x_3^2)}, \ \ A_3 = 0,
 \end{equation}
which satisfies
 \begin{equation}\label{RG}
 \partial_lA_m - \partial_mA_l = -\epsilon_{klm}\frac{x_k}{r^3} \quad \text{for} \;\; r\neq 0,
 \end{equation}
 as well as 
 \bee
 \vec{A}\cdot  d\vec{r} = \cos \beta \, d\alpha.
 \eee
 
  We now minimally couple the  motion on TN to the gauge potential (\ref{A}) via the Lagrangian
\begin{equation}
\label{minimalTN}
\mathcal{L}_p = \frac{V}{4}|\dot{\vec{r}}|^2 + \frac{L^2}{4V}(\dot{\gamma} + \vec{A}\cdot\dot{\vec{r}})^2 - \frac{p}{2V}(\dot{\gamma} +\vec{A}\cdot\dot{\vec{r}}).
\end{equation}
Clearly,   the  momentum $q$ conjugate to  the cyclic coordinate $\gamma$,
\begin{equation}
q = \frac{\partial\mathcal{L}_p}{\partial\dot{\gamma}} = \frac{L^2}{2V}(\dot{\gamma} + \vec{A}\cdot\dot{\vec{r}}) - \frac{p}{2V},
\end{equation}
is  conserved. The canonical momentum $\vec{\pi}$ conjugate to $\vec{r}$ is
\begin{align}
\label{pi2}
\vec{\pi} = \frac{\partial}{\partial\dot{\vec{r}}}\mathcal{L}_p = \frac{1}{2}V\dot{\vec{r}} + \frac{L^2}{2V}(\dot{\gamma} + \vec{A}\cdot\dot{\vec{r}})\vec{A} - \frac{p}{2V}\vec{A}= \vec{p} + q\vec{A}, 
\end{align}
where 
\bee
\vec{p} = \frac{1}{2}V\dot{\vec{r}}
\eee
 is called  the mechanical momentum \cite{GM}. 

The canonical symplectic structures on the phase space  $T^*M_{\text{\tiny TN}}$,
\bee
\label{TNsymplectic}
dx_l\wedge d\pi_l+ d\gamma\wedge dq,
\eee
is invariant under the $U(1)$ action which maps $\gamma \rightarrow \gamma +\delta$. The moment map for this action is the charge  $q$, viewed as map $T^*M_{\text{\tiny TN}}\rightarrow \RR$, and the symplectic quotient by this $U(1)$  action
\bee
\label{reducedphase}
\MM_q=T^*M_{\text{\tiny TN}}/\!\!/U(1)
\eee
is, by definition,   the pre-image of any real constant under the map $q$ divided by the $U(1)$ action.
 The position vector $\vec{r}$ and the canonical  momentum vector  $\vec{\pi}$ provide natural coordinates in terms of which the symplectic structure on $\MM_q$ takes the form 
\begin{align}
\label{TNsympl}
\omega = dx_l\wedge d\pi_l =
  dx_l\wedge dp_l + \frac{q}{2r^3}\epsilon_{iln}x_ndx_i\wedge dx_l.
\end{align}

The associated   Poisson brackets are
\begin{equation}
\label{Poisson}
\{A,B\} = \frac{\partial A}{\partial x_l}\frac{\partial B}{\partial\pi_l} - \frac{\partial A}{\partial \pi_l}\frac{\partial B}{\partial x_l},
\end{equation}
so that the mechanical  momentum $\vec{p} = \vec{\pi} - q\vec{A}$ satisfies
\begin{equation}\label{CR}
\{p_i,p_j\} = -q\epsilon_{ijk}\frac{x_k}{r^3}, \ \ \{p_i,f(\vec{r})\} = -\partial_i f(\vec{r}),
\end{equation}
where $f$ is any function  of $\vec{r}$.

We now rewrite the Lagrangian in terms of $\vec{p}$ and $q$,
\begin{equation}
\mathcal{L}_p = \frac{1}{V}|\vec{p}|^2 + \frac{q^2}{L^2}V - \frac{p^2}{4L^2V},
\end{equation}
and perform the Legendre transformation to obtain the gauged Hamiltonian, 
\begin{eqnarray}
\label{newH}
H_p &=& \dot{\vec{r}}\cdot\vec{\pi} + \dot{\gamma}q - \mathcal{L}_p \nonumber \\
&=& \frac{1}{V}|\vec{p}|^2 + \frac{q^2}{L^2}V + \frac{pq}{L^2} + \frac{p^2}{4L^2V} \nonumber \\
&=& H + \Delta H. 
\end{eqnarray}
Here $H$ is the  Hamiltonian  for $p=0$ and $\Delta H$ the contribution of the gauge potential:
\begin{equation}
H = \frac{1}{V}|\vec{p}|^2 + \frac{q^2}{L^2}V, \ \ \Delta H =  \frac{pq}{L^2} + \frac{p^2}{4L^2V}.
\end{equation}
Recalling that the  profile function $c$ appearing in the TN metric is $c=L/\sqrt{V}$, we note that 
\begin{align}
\label{eniq}
H_p & = \frac{1}{V}|\vec{p}|^2 + \left(\frac q c + \frac{pc}{2L^2}\right)^2  \nonumber \\
&  \geq  \left(\frac q c + \frac{pc}{2L^2}\right)^2 \nonumber \\
& \geq \begin{cases}  0 & \text{if} \; \;pq < 0 \\
2pq & \text{if} \; \;pq \geq 0. \end{cases}
\end{align}
For the last step we observed  that the second term in the first line is  the potential \eqref{simplepot} of the toy model of Sect.~2, and used the bound \eqref{toybound}. 

There is a conserved angular momentum \cite{GM} of $H$ given by   
\begin{equation}\label{diff}
\vec{J} = \vec{r}\times\vec{p} + q\hat{r},
\end{equation}
which, by  virtue of (\ref{CR}),  satisfies the relations
\begin{equation}\
\label{gJ}
\{J_k,p_l\} = \epsilon_{klm}p_m.
\end{equation}
It also follows that
\begin{eqnarray}
\label{JJ}
\{J_k,J_l\} = \epsilon_{klm}J_m.
\end{eqnarray}
Relation (\ref{gJ}) can be employed to check that $\vec{J}$ Poisson commutes with the Hamiltonian $H$. Since $\Delta H$ is spherically symmetric,  $\vec{J}$ also commutes with $H_p = H + \Delta H$.

In their study of  the geodesic motion on the negative mass TN space in \cite{GM},  Gibbons and Manton  showed  that there is  a  conserved  vector quantity analogous to the Runge-Lenz vector of the Kepler problem which takes the form
\begin{equation}\label{rl}
\vec{M} = \vec{p}\times\vec{J}  - \frac{ \hat{r}}{2L} \left(L^2H - 2q^2\right).
\end{equation}
One checks that it satisfies
\begin{equation}\label{JM}
\{J_k,M_l\} = \epsilon_{klm}M_m,
\end{equation}
and commutes with the TN Hamiltonian $H$ for any value (positive or negative) of $L$. However,  it fails to commute with our  gauged Hamiltonian $H_p$ since
\begin{equation}
\{\Delta H,M_k\} = -\frac{p^2}{4LrV^2}p_k + \frac{p^2}{4Lr^3V^3}x_k(\vec{r}\cdot\vec{p}).
\end{equation}
By trial and error we find that the vector-valued function 
\begin{equation}\label{f}
\vec{f} = \frac{p^2\vec{r}}{8LrV},
\end{equation}
satisfies $\{H_p, f_k\} = \{\Delta H,M_k\}$. Hence the components of the  gauged Runge-Lenz vector 
\begin{align}
\label{gM}
\vec{M}^p  = \vec{M} - \vec{f}  
= \vec{p}\times\vec{J} - \frac{\hat{r}}{2L} \left(L^2H_p - 2q^2 -pq\right),
\end{align}
commute with $H_p$. The Poisson brackets between the components of $\vec{J}$ and $\vec{M}^p$ turn out to be
\begin{align}
\label{poissonmj}
\{J_i,M^p_j\} &= \epsilon_{ijk}M^p_k,\nonumber \\
\{M^p_i,M^p_j\} &=  \left[\frac{1}{L^2}\left(q+\frac{p}{2}\right)^2 - H_p\right]\epsilon_{ijk}J_k.
\end{align}
We will study their Lie-algebraic interpretation in detail in Sects.~6 and 7.

\subsection{Classical trajectories}
The conserved quantities discussed above can be used to determine  the classical trajectories on TN in the gauged situation, i.e., the solutions of the Euler-Lagrange equations of \eqref{minimalTN} or Hamilton equation of \eqref{newH} with Poisson brackets \eqref{Poisson}. Considering first the simpler case where $q = 0$, we deduce from (\ref{diff}) and (\ref{gM})  that
\begin{equation}
\label{zeroqcond}
\vec{J}\cdot\hat{r} = 0, \ \  \vec{J}\cdot\vec{M}^p = 0, \ \ \vec{M}^p\cdot\vec{r} = J^2 -\frac{1}{2}LEr,
\end{equation}
where $E$ denotes the (constant) value of  $H_p$ and $J = |\vec{J}|$. The first and second equations show that the movement is in a plane orthogonal to $\vec{J}$ and that $\vec{M}^p$ is in this plane. Using polar coordinates $(r,\phi)$ to parametrise the plane with $\vec{M}^p$ in the direction determined by $\phi = 0$, we deduce, from the third equation,
\begin{equation}
r = \frac{J^2}{|\vec{M}^p|\cos\phi + \frac{1}{2}LE}.
\end{equation}
This is the equation of a conic section. Finally taking into account the relation
\begin{equation}
\label{cone1}
|\vec{M}^p| = \sqrt{\left(E - \frac{p^2}{4L^2}\right)J^2 + \frac{1}{4}L^2E^2},
\end{equation}
we obtain the following types of orbit:
  the conic section is an ellipse for $L^2E<\frac{p^2}{4}$, a parabola for $L^2E = \frac{p^2}{4}$ and a hyperbola for $L^2E > \frac{p^2}{4} $.

\begin{figure}[!h]
\centering
\includegraphics[width=8truecm]{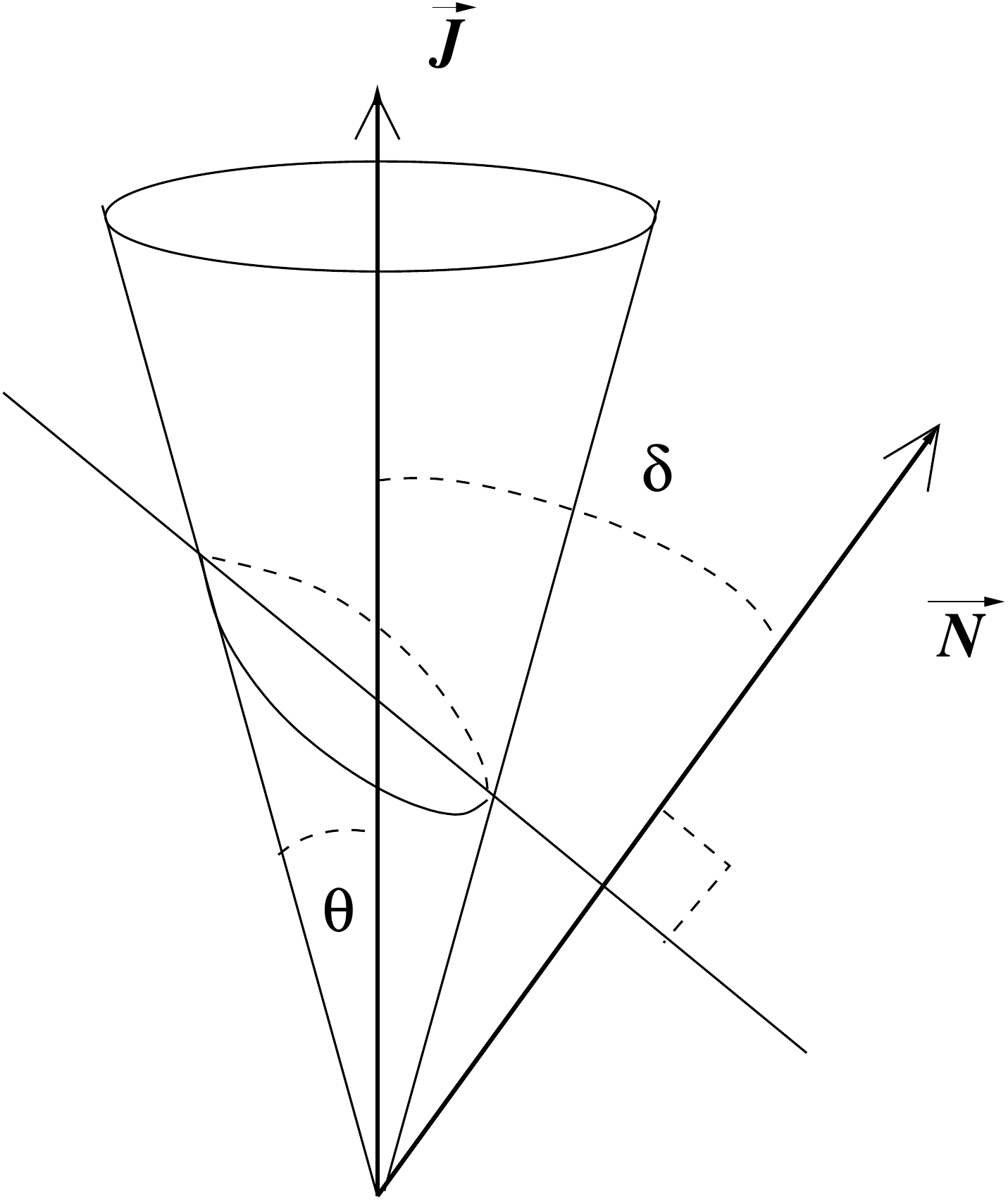}
\caption{The conic sections determined by the conserved  vectors $\vec{J}$ and $\vec{N}$.}
\label{conicsections}
\end{figure}

In the  general case $q\neq 0$,  the expression (\ref{diff})   implies
\begin{equation}
\label{coneq}
\vec{J}\cdot\hat{r} = q, 
\end{equation}
which  shows that $\vec{r}$ lies on a cone whose axis of symmetry is along $\vec{J}$ and  whose vertex is at the origin. The opening angle $2\theta\in (0,\pi)$ of the cone relative to direction of $\vec{J}$ as shown in Fig.~\ref{conicsections} is determined by 
\bee
\label{opendef}
\cos \theta = \frac{|q|}{J}.
\eee
For $q >0$, the  cone is in the `positive'   half-space determined by $\vec{J}\cdot \vec{r} >0$, while  $q < 0$ it is in the `negative'  half-space determined by $\vec{J}\cdot \vec{r} <0$.

Furthermore, the equations  (\ref{diff})  and \eqref{gM} imply
\begin{equation}
\label{mastereq}
\vec{J}\cdot\vec{M}^p = -\frac{q}{2L}\left(L^2E - 2q^2-pq\right),  \quad  \vec{M}^p\cdot\vec{r} = J^2-q^2 -\frac{r}{2L}\left(L^2E - 2q^2 -pq\right).
\end{equation}
To interpret them, we define the vector
\begin{equation}
\vec{N} = q\vec{M}^p + \frac{1}{2L}\left(L^2E -2q^2 -pq \right)\vec{J}.
\end{equation}
As a linear combination of conserved vectors with conserved coefficients, this vector is also conserved.  In terms of this vector, the second  equation in \eqref{mastereq} 
is equivalent to 
\bee
\vec{N}\cdot\vec{r} = q(J^2-q^2),
\eee 
which shows that the  motion is also in a plane perpendicular to the vector $\vec{N}$. With the notation  $l = |\vec{r}\times\vec{p}|$ for the magnitude of the orbital angular momentum, we note 
\bee
\label{jlq}
J^2 = l^2 + q^2,
\eee
so that 
\begin{equation}
\label{nplane}
\vec{N}\cdot\vec{r} = ql^2.
\end{equation}

The classical trajectories in the case $q\neq 0$ are thus intersections of the cone defined by \eqref{coneq} and the plane defined by \eqref{nplane}.  From classical geometry we know that these are ellipses (including the degenerate case of a point), parabolae or  hyperbolae (including the degenerate case of a line). The nature of the orbit depends on the energy $E$ and on the relative size of $q$ and $p$; as we shall see, the details are quite  subtle, combining  the results from the toy model in Sect.~2 with lessons from the role of conic sections as trajectories in the standard Kepler problem. 

Focusing  on the non-degenerate case $l\neq 0$, we note that the sign of $q$ determines both the direction of the cone  \eqref{coneq} and the position of the plane \eqref{nplane} relative to the origin. If $q>0$ then the situation is as shown in Fig.~\ref{conicsections}, with the  cone in the positive  half-space determined by $\vec{J}\cdot \vec{r} >0$ and the plane \eqref{nplane} displaced from the origin in the direction of $\vec{N}$. If $q<0$ the cone is in the opposite half-space and the plane is displaced from the origin in the direction of $-\vec{N}$. The nature of the intersection between them, however, is independent of the sign of $q$, and only depends on the angle between $\vec{J}$ and $\vec{N}$, see again Fig.~\ref{conicsections}.

A  lengthy calculation shows that the 
 squared norm  of $\vec{N}$ is 
\bee
|\vec{N}|^2= \frac{ l^2 E }{4}\left(L^2E - {2pq}\right),
\eee 
which is positive for all allowed values of the energy by virtue of \eqref{eniq}.
Since, from the first equation in \eqref{mastereq}, 
\begin{equation}
 \vec{N}\cdot\vec{J} = \frac{l^2}{2L}(L^2E - 2q^2 - pq),
\end{equation}
 we deduce  that $\delta$ is determined  by
\begin{equation}
\label{deltadef}
\cos{\delta} = \frac{l}{J}\frac{L^2E -2q^2  -pq}{L\sqrt{E(L^2E -2pq)}  }.
\end{equation}
 In order to classify the orbits we also note that,   from \eqref{opendef} and \eqref{jlq},    $\sin{\theta} = \frac{l}{J}$ or 
\begin{equation}
\label{thetadef}
\cos\left( \frac{\pi}{2} - \theta\right) = \frac{l}{J}.
\end{equation}

Elementary geometrical considerations in Fig.~\ref{conicsections} now show that 
\begin{equation}
\label{orbitcond}
l\neq 0 \quad \text{and} \quad  \left. \begin{cases} 
\delta < \frac{\pi}{2} -\theta \\  
\delta = \frac{\pi}{2} -\theta   \\ 
 \frac{\pi}{2} -\theta < \delta < \frac{\pi}{2} +\theta \\
 \delta \geq  \frac{\pi}{2} +\theta
  \end{cases}\right \}
 \quad  \Leftrightarrow \quad  \text{orbit is}  \quad  \left. \begin{cases} \text{ellipse} \\ \text{parabola}  \\ \text{hyperbola}  \\ \text{empty set}.
 \end{cases}\right \}
\end{equation}

We analyse each of those conditions in turn. Since  the cosine function  is strictly decreasing on the interval $[0,\pi]$,  applying it to the inequalities in \eqref{orbitcond} reverses them.  It will also be useful to observe that the energy bound  \eqref{eniq}  implies
\bee
\label{neat}
q^2 <\frac{p^2}{4} \Rightarrow L^2E>2q^2+ pq.
\eee
For the short proof, one needs to distinguish the cases $pq>0$ and $pq<0$ and use $|q|<|p/2|$.

For elliptic orbits, we  require  $\cos{\delta}>\cos(\frac{\pi}{2} - \theta)$. Inserting the above relations,  this condition gives 
\begin{equation}
\label{condell}
 l \neq 0 \quad \text{and} \quad  L^2E -  2 q^2 - pq > L\sqrt{E(L^2 E - 2pq)}.
\end{equation}
Since the right hand side is positive (assuming $L>0$), we deduce that, for elliptic orbits, 
\bee
\label{Epqclassical}
 L^2E > 2q^2+ pq.
\eee
On the other hand, squaring both sides of   \eqref{condell}, we deduce
\bee
\label{ellipsecond}
 L^2E  <  \left(q+\frac{p}{2}\right)^2.
\eee
However, the inequalities \eqref{Epqclassical} and \eqref{ellipsecond} can only both be satisfied if 
\bee
\label{pqcond}
 q^2 <\frac{p^2}{4},
\eee
which is precisely the condition \eqref{toycond} derived in the toy model in Sect.~2.
Since, by \eqref{neat},  the condition \eqref{pqcond} is sufficient for \eqref{Epqclassical} to hold, we deduce that elliptical orbits occur iff  $p\neq 0$, the  charge $q$ satisfy \eqref{pqcond} and the energy satisfies \eqref{ellipsecond} \footnote{ If $p$ were to vanish then \eqref{pqcond} forces $q$ to vanish, and then \eqref{ellipsecond} would imply $E=0$, which is impossible.}. As an aside we note that elliptical orbits  are possible in the case  $L<0$ even when $p=0$ (as discussed in \cite{GM}). 

Returning to general $p$  and positive $L$,  the  analysis of the  conditions \eqref{orbitcond} for the parabolic and  hyperbolic cases along the lines of the discussion of elliptical orbits is now straightforward. We skip most details, but point out that, in the hyperbolic case, the trigonometric identity $\cos\left(\frac \pi 2 +\theta\right) = -\cos\left(\frac \pi 2 -\theta\right) $  applied to \eqref{orbitcond} implies the condition 
\begin{equation}
\label{condhyp}
 l \neq 0 \quad \text{and} \quad  |L^2E -  2 q^2 - pq | <   L \sqrt{E(L^2 E - 2pq)},
\end{equation}
which (for positive $L$)  is equivalent to 
\bee
\label{hyperbolic}
 L^2E  >  \left(q+\frac{p}{2}\right)^2,
\eee
but does not require any restrictions on $p$ and $q$. 

We summarise the  dependence of the orbits on the energy $E$ and the charge $q$ as follows:
\begin{equation}
\label{orbits}
l\neq 0 \;\; \text{and} \;\;  \left. \begin{cases}
p\neq 0, \;\;q^2<\frac{p^2}{4}, \;\;  L^2E  <  \left(q+\frac{p}{2}\right)^2\\
p\neq 0,\; \;q^2\leq \frac{p^2}{4}, \;\; L^2E  = \left(q+\frac{p}{2}\right)^2 
\\  L^2E >  \left(q+\frac{p}{2}\right)^2 \end{cases}\right \}
 \;\;  \Leftrightarrow \;\;  \text{orbit is}  \;\;  \left. \begin{cases} \text{ellipse} \\ \text{parabola}  \\ \text{hyperbola}.
 \end{cases}\right \}
\end{equation}

 \section{Gauged Taub-NUT quantum mechanics } 
\subsection{Canonical quantisation} 
In  this paper we set $\hbar =1$ when discussing quantum mechanics. With this convention, 
the canonical quantisation procedure  of $T^*M_{\text{\tiny TN}}$ amounts to replacing
\begin{equation}
\vec{\pi} \to -i\frac{\partial}{\partial \vec{r}}, \ \ q \to -i\partial_{\gamma},
\end{equation}
where $\frac{\partial}{\partial \vec{r}}=\left(\frac{\partial}{\partial x_1}, \frac{\partial} {\partial x_2},\frac{\partial}{\partial x_3}\right)$.
A comparison of $q$ with \eqref{dual} shows that, as an operator, 
\bee
\label{qx3} 
q = -i\partial_{\gamma} = -iX_3.
\eee
 The  relation \eqref{pi2} implies the quantisation of the mechanical momentum according to
 \bee
 \vec{p} \to -i\frac{\partial}{\partial \vec{r}} +i\vec{A}\partial_\gamma,
 \eee
where $\vec{A}$ is the magnetic monopole vector potential \eqref{Dmonopole}.

 Inserting \eqref{qx3}  into \eqref{newH} gives
\begin{eqnarray}\label{prove}
H_p
&=& \frac{1}{V}|\vec{p}|^2 - \frac{V}{L^2}\left(X_3 + \frac{ip}{2V}\right)^2,
\end{eqnarray}
which turns out to be precisely the gauged Laplace operator \eqref{Hp}.
To see this,  note that 
\begin{eqnarray}
|\vec{p}|^2 &=& (-i\partial_l + i\partial_{\gamma}A_l)(-i\partial_l + i\partial_{\gamma}A_l)\nonumber \\
&=& -\left(\frac{\partial}{\partial \vec{r}}\right)^2 + 2\left(\vec{A}\cdot \frac{\partial}{\partial \vec{r}}\right)\partial_{\gamma}  - \vert\vec{A}\vert^2 \partial_{\gamma}^2,
\end{eqnarray}
where $\left(\frac{\partial}{\partial \vec{r}}\right)^2$ is the Laplace operator on Euclidean $\RR^3$, and  we have used that,  for the Dirac monopole \eqref{Dmonopole}, $ \text{div}\vec{A} = 0$. In terms of  spherical coordinates and \eqref{Dmonopole} one checks that 
\begin{equation}
\left(\frac{\partial}{\partial \vec{r}}\right)^2 = \frac{1}{r^2}\partial_r(r^2\partial_r) + \frac{1}{r^2}(\partial_{\beta}^2 + \cot{\beta}\partial_{\beta} +\csc^2{\beta}\partial_{\alpha}^2), \quad 
\vec{A}\cdot \frac{\partial}{\partial \vec{r}} = \frac{\cos{\beta}}{r^2\sin^2{\beta}}\partial_{\alpha},
\quad 
\vert\vec{A}\vert^2 = \frac{\cos^2{\beta}}{r^2\sin^2{\beta}}, 
\end{equation}
so that  
\begin{align}\label{X2}
|\vec{p}|^2 &= -\frac{1}{r^2}\partial_r(r^2\partial_r) - \frac{1}{r^2}(\partial_{\beta}^2 + \cot{\beta}\partial_{\beta} +\csc^2{\beta}\partial_{\alpha}^2 - 2\cot{\beta}\csc{\beta}\partial_{\gamma}\partial_{\alpha} + \cot^2{\beta}\partial_{\gamma}^2) \nonumber \\
&= -\frac{1}{r^2}\partial_r(r^2\partial_r) - \frac{1}{r^2}(X_1^2+X_2^2),
\end{align}
where we have used the relation,
\begin{equation}
\label{X1X2}
X_1^2 + X_2^2 = \partial_{\beta}^2 + \cot{\beta}\partial_{\beta} + \cot^2{\beta}\partial_{\gamma}^2 + \csc^2{\beta}\partial_{\alpha}^2 - 2\cot{\beta}\csc{\beta}\partial_{\alpha}\partial_{\gamma},
\end{equation}
which can be obtained from \eqref{dual}. Substituting  \eqref{X2} into \eqref{prove} shows that the quantum Hamiltonian $H_p$ is the gauged Laplace operator \eqref{Hp}, as claimed.


Finally applying the quantisation rule to the angular momentum  $\vec{J}$ defined in \eqref{diff} we obtain the differential operator
\begin{align}
\label{quantumJ}
\vec{J}&=-i\vec{r}\times \frac{\partial}{\partial \vec{r}} + i(\vec{r}\times \vec{A} -\hat{r} )\partial_\gamma  \nonumber \\
& =i\begin{pmatrix} \sin\alpha \, \partial_\beta +\cot\beta\cos \alpha\,  \partial_\alpha -\frac{\cos\alpha}{\sin \beta} \partial_\gamma \\-\cos \alpha \, \partial_\beta +\cot\beta \sin \alpha \, \partial_\alpha -\frac{\sin\alpha}{\sin \beta} \partial_\gamma \\ 
-\partial_\alpha
\end{pmatrix}.
\end{align}
Transforming coordinates according to 
\eqref{s3},  one checks that, up a factor of $i$, the components are the vector fields $Z_1,Z_2$ and $Z_3$ \eqref{spacefixed}  generating the left-action of $SU(2)$ on itself:
\bee
\vec{J} = i \vec{Z}.
\eee
It follows that the squared total angular momentum operator can be written in terms of the left- and right-generated vector fields on $S^3$ as 
\bee
\label{totalJ}
\vec{J}^2 = -(Z_1^2+Z_2^2 +Z_3^2) = -(X_1^2+X_2^2 +X_3^2) = -\Delta_{S^3}.
\eee

\subsection{Separating variables}

 For fixed $r$, the angular part of the  quantum Hamiltonian \eqref{Hp} 
  is akin to the Hamiltonian of  a symmetric rigid body coupled to a gauge field. In that context, the   operators $iZ_j$  are interpreted as `space-fixed' angular momentum  components and the operators $iX_j$ as `body-fixed' angular momentum  components \cite{GM,JS}. The quantum Hamiltonian $H_p$  commutes with $Z_1,Z_2,Z_3$ and with $X_3$; together,  these generate the $U(2)$ symmetry of TN space. 

To separate the radial from the angular dependence in the wavefunction, we therefore require  a complete set of functions on $SU(2)$ which diagonalise the commuting operators $\Delta_{S^3}, iZ_3,iX_3$. This is usually done in terms of Wigner functions of the Euler angles, but here we use the construction of  the  eigenfunctions as homogeneous polynomials of the complex coordinates $z_1,z_2$ and their complex conjugates given in \cite{JS}. As explained there, an irreducible representation of $SU(2)$ can be given in terms of polynomials in $z_1,z_2, \bar{z}_1,\bar{z}_2$ that belong to the kernel of the differential operator $\Box = 4(\partial_1\bar{\partial}_1 + \partial_2\bar{\partial}_2)$. Combining this observation with the discussion in \cite{Dray2} we obtain a basis that satisfies this irreducibility condition:
\begin{equation}
\label{Y}
Y^j_{sm} = \left[\frac{(j+s)!(j-s)!}{(j+m)!(j-m)!}\right]^{1/2}\sum_k\frac{(j+m)!}{(j+m-k)!k!}\frac{(j-m)!(-1)^{j-s-k}}{(j-s -k)!(s -m + k)!}z_1^{s-m+k}z_2^{j+m-k}\bar{z}_1^k\bar{z}_2^{j-s-k},
\end{equation}
where 
\bee
j\in\frac 1 2 \ZZ^+, \quad s, m=-j,-j+1,\ldots,j-1,j, 
\eee
and  $k$ runs over the values so that the factorials are well defined.
These functions are normalised and are clearly orthogonal since they are eigenfunctions of the Hermitian operators $\Delta_{S^3},$ (total angular momentum), $ iZ_3$ (angular momentum along the space-fixed 3-axis), $iX_3$  (angular momentum along the body-fixed 3-axis) with eigenvalues
\begin{equation}
\label{FI}
\Delta_{S^3}Y^j_{sm} = -j(j+1)Y^j_{sm}, \ \ iZ_3Y^j_{sm} = mY^j_{sm}, \ \ iX_3Y^j_{sm} = sY^j_{sm}.
\end{equation}
They also satisfy 
\begin{equation}
X_+Y^j_{sm} = -i[(j-s)(j+s+1)]^{1/2}Y^j_{s+1,m}, \ \ X_-Y^j_{sm} = -i[(j+s)(j-s+1)]^{1/2}Y^j_{s-1,m},
\end{equation}
where $X_{\pm} = X_1 \pm iX_2$, which shows that all the angular momentum eigenstates can be obtained from the holomorphic  $Y^j_{jm}$  or the anti-holomorphic $Y^j_{-jm}$  by the repeated action of $X_-$  or $X_+$.

We look for stationary states $\Psi$ 
of the form 
\bee
\Psi(r,z_1,z_2) = R(r)Y^j_{sm}(z_1,z_2).
\eee
Using
\begin{equation}
\label{W}
 (X_1^2 + X_2^2 )Y^j_{sm} = [-j(j+1) + s^2]Y^j_{sm}
\end{equation}
in the stationary Schr\"odinger equation
\begin{equation}
\label{eigen}
H_p\Psi = E\Psi,
\end{equation}
we obtain  the radial equation
\begin{equation}
\label{radialSch}
\left[-\frac{1}{r^2}\partial_r(r^2\partial_r) + \frac{1}{r^2}j(j+1) + \left(\frac{2s^2}{L} - \frac{ps}{L} - EL\right)\frac{1}{r} + \left(\left(\frac{s-\frac{p}{2}}{L}\right)^2 - E\right)\right]R(r) = 0.
\end{equation}
 Before we study  bound and scattering states in the  following sections, we make two general observations.

It follows from  \eqref{qx3} that, when acting on the functions \eqref{Y}, the operator $q$ has the eigenvalue $-s$. For later use, note that the classical bound \eqref{eniq} also holds in the quantum case, so that, in particular, for any eigenstate of  $H_p$  and $q$ with eigenvalues $E$ and $-s$, we have 
\bee
\label{quanteniq}
L^2E  \geq \begin{cases} \phantom{-} 0 & \text{if} \; ps > 0 \\
-2sp & \text{if} \; ps  \leq 0. \end{cases}
\eee

Finally, It is worth stressing that neither the space-fixed nor the body-fixed angular momentum operators discussed above are  invariant under $U(1)$-gauge transformations. The quantum numbers $j,s$ and $m$ are not gauge invariant either and therefore have to be interpreted with care. However, this is familiar in the context of the 
Schr\"odinger  equation coupled to a magnetic field, particularly in the discussion of Landau levels for planar motion in a magnetic field. Even though the angular momentum operator is not gauge invariant in this context, the eigenvalues can be used to label degenerate energy eigenstates. This labelling is not gauge invariant, but physical quantities like the energy or the degeneracy of energy levels are. The role of the gauge choice in labelling degenerate states in Landau levels is discussed in detail in \cite{HRR}, see also the book \cite{Fradkin}.

 \subsection{Bound states}
The substitution of
\begin{equation}
\label{kprimed}
R(r) = r^je^{-k'r}u(r), \ \  k'^2 = \left(\frac{s-\frac{p}{2}}{L}\right)^2 - E,
\end{equation}
into  the radial Schr\"odinger equation \eqref{radialSch},  
reduces it to 
\begin{equation}
\label{hgeq}
z\frac{d^2u}{dz^2} + (b- z)\frac{du}{dz} - a u(z) = 0,
\end{equation}
where
\begin{equation}
z = 2k'r, \ \ a  = (j+1 + \lambda), \ b = 2j +2, 
\end{equation}
and 
\begin{equation}
\label{lambdaformula}
\lambda = -\frac{1}{2k'L}\left(L^2E + ps - 2s^2\right).
\end{equation}

The equation \eqref{hgeq}  is the confluent hypergeometric equation  \cite{AS}. The general solution which is regular at the origin is 
\begin{equation}
u = Ar^je^{-k'r}M(a,b,z),
\end{equation}
where $A$ is an arbitrary constant and $M$ is Kummer's function of the first kind. 
Square integrability  requires 
\bee 
a = -\nu, \qquad \nu = 0,1,2,\ldots  \quad \text{and} \quad k' \in \RR^+.
\eee
Since $j$ takes arbitrary half-integer positive values,  the first condition is equivalent to 
 \bee
 \label{lambdaquant}
 n:=-\lambda =  \nu+ j+1, \qquad \nu = 0,1,2,\ldots.
 \eee
In principle, $n$ can take all  half-integer values $\geq 1$, but  the ranges of the  quantum numbers $j$ and $n$ are related by 
\bee
\label{jnrel}
n=j+1+ \nu,  \qquad \nu = 0,1,2,\ldots .
\eee

This requirement together with  the expression  \eqref{lambdaformula} for $\lambda$ as well as $L>0$ and $k'>0$  imply
\begin{equation}
\label{Ebigger}
L^2E >  2s^2-  ps.
\end{equation}
On the other hand, 
  the relation \eqref{kprimed}  between $E$ and $k'$  enforces
\begin{equation}
\label{SC}
L^2E <  \left(s-\frac{p}{2}\right)^2= s^2 - ps + \frac{p^2}{4}.
\end{equation}
There can only be bound states if these two inequalities can be simultaneously satisfied, i.e.,  if
\begin{equation}
\label{bdcond}
\frac{p^2}{4} > s^2 ,
\end{equation}
which is the quantum version of the  condition \eqref{pqcond} for bounded orbits in the classical theory.

Note that, if $L$ were negative, the inequality \eqref{Ebigger} would have the opposite direction and there would be no condition on $p$.  In that case we can set $p=0$ and recover the bound states in the  singular $L=-2$  TN space discussed in \cite{GM}, which exist for any $s\neq 0$. As  shown   in \cite{dVS}, there are no bound states (and no bounded orbits) when $L>0$ and  $p=0$. More generally, however,   binding is always possible when $p$ is sufficiently large. All these results confirm the qualitative discussion of the two-dimensional  toy model  in Sect.~\ref{toysect}.  

Solving \eqref{kprimed} and \eqref{lambdaquant} for $E$, we  find 
\begin{equation}
\label{SE}
E = \frac{2}{L^2}\left[-n^2 + s^2 - \frac{ps}{2}\right] \pm \frac{2n}{L^2}\sqrt{n^2 - s^2 + \frac{p^2}{4}}.
\end{equation}
Only the solution with the upper sign satisfies \eqref{Ebigger}, and we write the resulting spectrum of bound state energies  as 
\bee
\label{Espectrum}
E = \frac{2}{L^2}\left[  s^2 - \frac{ ps}{2}  +  n\sqrt{n^2 - s^2+\frac{p^2}{4}} - n^2\right],
\qquad n  =  |s|+1, |s|+2,  |s|+3 \ldots.
\eee
The behaviour for large $n$ is typical for Coulomb bound states
\bee
E\approx \frac{\left(s-\frac p 2 \right)^2} {L^2}  - \frac {\left(s^2-\frac{p^2}{4}\right)^2 }{4L^2 n^2} + \mathcal{O}\left(\frac{1}{n^4}\right).
\eee

This formula for the bound state energy shows that, like in the toy model of Sect.~2,  the  bound state energies are relatively  high when $p$ and $q$ have the same sign (so that  the signs of  $p$ and $s$  are opposite)   but are lowered when the signs  of $p$ and $q$ are opposite (and those of $p$ and $s$ the same). Note  also that, in the limit $p=0$ and for $L=-2$ our formula reduces to that  obtained in \cite{GM} for the negative mass TN space. For a detailed comparison observe that in \cite{GM} only integer values of $j$ and $n$ were considered.

The energy levels  for fixed $s$ and $n$ have a large degeneracy, given by the sum over the dimension $2j+1$ for allowed values of $j$. Recalling the constraint \eqref{jnrel} and  $j\geq |s|$, the degeneracy is 
\begin{equation}
\label{degeneracy}
{\sum}_{j=|s|}^{'n-1}2j+1= (2|s|+1)+ (2|s|+3) + \ldots + (2n-3)+ (2n-1) = n^2-s^2,
\end{equation}
with $\sum'$ indicating that we sum over integers if $|s|$ is an integer and over half-odd integers if $|s|$ is a half-odd integer. As we shall see in Sect.~6, the degeneracy can be understood in terms of a  conserved Runge-Lenz vector.

\subsection{Scattering states}

Next we turn to solutions of  the eigenvalue equation \eqref{eigen}  which describe stationary scattering states.
For the analysis of scattering it is convenient to use 
parabolic coordinates familiar from the treatment of  Coulomb scattering. 

 Assuming solutions of \eqref{eigen} of the form 
\bee
\label{scattfun}
\Psi = e^{-is\gamma}e^{im\alpha}\Lambda(\beta,r),
\eee 
and recalling the formula \eqref{X1X2} and
we  find that $\Lambda$ has to satisfy the equation
\begin{align}
\label{Lambdawave}
\left(\partial_r^2 + \frac{2}{r}\partial_r+ \frac{1}{r^2}\partial_{\beta}^2 + \frac{1}{r^2}\cot{\beta}\partial_{\beta}\right)\Lambda + \frac{1}{r^2}(-s^2\cot^2{\beta} - m^2\csc^2{\beta} + 2ms\cot{\beta}\csc{\beta})\Lambda \nonumber \\
- \frac{V^2}{L^2}\left(s - \frac{p}{2V}\right)^2\Lambda + EV\Lambda = 0. 
\end{align}
Now introducing parabolic coordinates $\xi, \eta$ via
\begin{equation}
\label{xieta}
\xi = r(1 + \cos{\beta}),  \qquad 
\eta = r(1 - \cos{\beta}),
\end{equation}
and noting the inverse transformation
\begin{equation}
\label{xietainv}
r = \frac{\xi + \eta}{2}, \ \ \cos\beta = \frac{\xi - \eta}{\xi + \eta}, \ \ \sin{\beta} = \frac{2\sqrt{\xi\eta}}{\xi + \eta}, 
\end{equation}
we think of $\Lambda$ now as a function of $\xi$ and $\eta$ via the substitution \eqref{xietainv} for $r$ and $\beta$. 
Then \eqref{Lambdawave}  becomes 
\begin{eqnarray}
\frac{4}{\xi + \eta}\left(\xi\partial_{\xi}^2 + \eta\partial_{\eta}^2 + \partial_{\xi} + \partial_{\eta}\right)\Lambda - \frac{1}{\xi\eta}\left(s^2 + m^2 + 2ms\frac{\xi-\eta}{\xi+\eta}\right)\Lambda \nonumber \\
+ \left(-\frac{2s^2}{L} + \frac{sp}{L} + EL\right)\frac{2\Lambda}{\xi + \eta}  - \frac{1}{L^2}\left(s - \frac{p}{2}\right)^2\Lambda + E\Lambda = 0. 
\end{eqnarray}
Separating variables again via  $\Lambda = f(\xi)g(\eta)$ we find
\begin{equation}
\frac{4\xi}{f}\partial_{\xi}^2f + \frac{4}{f}\partial_{\xi}f - \frac{1}{\xi}(m + s)^2 + k^2\xi + 2\left(EL - \frac{2s^2}{L} + \frac{ps}{L}\right) - C = 0, 
\end{equation}
\begin{equation}
\frac{4\eta}{g}\partial_{\eta}^2g + \frac{4}{g}\partial_{\eta}g - \frac{1}{\eta}(m - s)^2 + k^2\eta + C = 0,
\end{equation}
where  $C$  is a separation constant and
\begin{equation}
k^2 = E - \frac{1}{L^2}\left(s - \frac{p}{2}\right)^2.
\end{equation}
These differential equations can be simplified further if we assume solutions of the form
\begin{equation}
f(\xi) = \xi^{\frac{|m-s|}{2}}e^{-\frac{ik\xi}{2}}F(\xi), \ \  g(\eta) = \eta^{\frac{|m+s|}{2}}e^{-\frac{ik\eta}{2}}G(\eta), 
\end{equation}
and implement the change of variable
\begin{equation}
z_1 = ik\xi, \ \ \ z_2 = ik\eta.
\end{equation}
Doing so we deduce that both $F$ and $G$ satisfy the confluent hypergeometric equation
\begin{equation}
z_1\frac{d^2F}{dz_1^2} + (b_1 - z_1)\frac{dF}{dz_1} - a_1F = 0,
\end{equation}
\begin{equation}
z_2\frac{d^2G}{dz_2^2} + (b_2 - z_2)\frac{dG}{dz_1} - a_2G = 0,
\end{equation}
where
\begin{equation}
a_1 = \frac{|m-s|}{2} + \frac{1}{2} - \frac{ic}{4k} + \frac{i}{2k}\left(EL - \frac{2s^2}{L} + \frac{ps}{L}\right), \ \ b_1 = |m-s| + 1,
\end{equation}
\begin{equation}
a_2 = \frac{|m+s|}{2} + \frac{1}{2} + \frac{ic}{4k}, \ \ b_2 = |m+s| + 1.
\end{equation}
We see from these relations that
\begin{equation}
a_1 + a_2 = 1 + \frac{1}{2}|m+s| + \frac{1}{2}|m-s| - i\lambda,
\end{equation}
where
\begin{equation}
\label{lambdascatt}
\lambda = -\frac{L}{2kL^2}\left(L^2E +ps - 2s^2 \right).
\end{equation}
So, the scattering solution is of the form
\begin{equation}
\Psi = e^{-is\gamma}e^{im\alpha}\xi^{\frac{|m-s|}{2}}\eta^{\frac{|m+s|}{2}}e^{-\frac{ik\xi}{2}}e^{-\frac{ik\eta}{2}}M(a_1,b_1,ik\xi)M(a_2,b_2,ik\eta),
\end{equation}
where $M$ is Kummer's function of the first kind. This is formally the same as the  two monopole scattering state found by Gibbons and Manton \cite{GM} in the case $p=0$ and $L = -2$. However, in our case the constants $k$ and $\lambda$ have an extra $p$-dependence and  the length parameter $L$ is positive. 

As in \cite{GM} one can compute the cross section by looking at the wave function with $m=s$ and $a_1 = 1$,
\begin{equation}
\Psi = e^{is(\alpha - \gamma)}(r - z)^{|s|}M(|s|-i\lambda, 2|s|+1, ik(r-z)), \ \ z = r\cos{\beta}.
\end{equation}
Then the substitution of the asymptotic form of $M(|s|-i\lambda, 2|s|+1, ik(r-z))$ for large $|z|$ as in \cite{GM} allows us to identify the scattered spherical wave, and to obtain the cross  section
\begin{equation}
\frac{d\sigma}{d\Omega} = \frac{s^2 + \lambda^2}{4k^2}\csc^4{\textstyle\frac{\beta}{2}}.
\end{equation}
Writing $\lambda$ in terms of $k$ via \eqref{lambdascatt},
we finally arrive at the  cross section 
\begin{eqnarray}
\label{pscross}
\left(\frac{d\sigma}{d\Omega} \right)_{(p,s)}&=& \frac{L^2}{16}\left[\frac{4s^2}{k^2L^2} + \left(1 - \frac{s^2}{k^2L^2}  + \frac{p^2}{4k^2L^2}  \right)^2\right]\csc^4{\textstyle\frac{\beta}{2}} \nonumber \\
&=& \frac{L^2}{16}\left[\left(1 + \frac{s^2}{k^2L^2}\right)^2 + \frac{p^2}{4k^2L^2}\left(2 - \frac{2s^2}{k^2L^2} + \frac{p^2}{4k^2L^2}\right)\right]\csc^4{\textstyle\frac{\beta}{2}}. 
\end{eqnarray}

The special cases $s=0$  and $p=0$ are interesting because the resulting cross sections
\bee
\label{magscat}
\left(\frac{d\sigma}{d\Omega}\right)_{(p,0)} = \frac{L^2}{16} \left(1+ \frac{(p/2)^2}{L^2k^2}\right)^2\csc^4{\textstyle\frac{\beta}{2}}
\eee
and 
\bee
\label{elecscat}
\left(\frac{d\sigma}{d\Omega}\right)_{(0,s)} = \frac{L^2}{16} \left(1+ \frac{s^2}{L^2k^2}\right)^2\csc^4{\textstyle\frac{\beta}{2}}
\eee
are mapped into each other under  the exchange $s \leftrightarrow p/2$ even though the general case  \eqref{pscross} is not invariant under this exchange. 
In both special cases,  the charge,  energy and angular  dependence approaches that of the  Rutherford scattering cross section for electrically charged particles in the limit of large $s$ (or $p$). 

\section{Algebraic calculation of  quantum bound states}

\subsection{The  Runge-Lenz operator and $so(4)$ symmetry} 

In 1926, Pauli  computed the quantum spectrum of the Kepler problem by using the conservation of the Runge-Lenz vector \cite{Pauli}. His method has since then been much    explored  and extended in various papers,  see  \cite{BI}  for a reference which is particularly useful in  the current context. More recently, it  was used to compute  bound states and scattering or the Laplace operator on TN space \cite{FH} and also for the Dirac operator on TN \cite{CH}. We now use it to re-derive the spectrum of the gauged TN  Hamiltonian \eqref{Hp} purely algebraically.

As always in quantising a theory, we need to be careful with ordering  in the quantisation of classically conserved quantities. While there are no such ambiguities in the definition of the angular momentum operators, they do arise in defining a quantum version of the  Runge-Lenz vector. The quantum analogues  of the canonical Poisson brackets (\ref{CR}),
\begin{equation}
[p_i,p_j] = -iq\epsilon_{ijk}\frac{x_k}{r^3}, \ \ [p_j,f(\vec{r})] = -i\partial_jf(\vec{r}),
\end{equation}
imply, for the quantum angular momentum operator \eqref{quantumJ}, 
\begin{equation}
\label{Jp}
[J_i,p_j] = i\epsilon_{ijk}p_k,
\end{equation}
which is the quantum version of \eqref{gJ}. This means that $[J_i,p_j]\neq 0, \ i\neq j$ and hence the order of $\vec{J}$ and $\vec{p}$ is important in the definition of the quantum version of the Runge-Lenz vector \eqref{gM}.

Noting that, classically,  $\vec{p}\times\vec{J}= \frac{1}{2}(\vec{p}\times\vec{J} - \vec{J}\times\vec{p})$,  one finds  that the quantum ordering
\begin{equation}
\vec{M} = \frac{1}{2}(\vec{p}\times\vec{J} - \vec{J}\times\vec{p}) -\frac{\hat{r}}{2L}\left(L^2H - 2q^2\right)
\end{equation}
ensures  that the  quantum commutation relations between $\vec{J}$ and $\vec{M}$ are,  
\begin{eqnarray}\label{QR}
[J_k,J_l] = i\epsilon_{klm}J_m, \ \ [J_k,M_l] = i\epsilon_{klm}M_m,
\end{eqnarray}
in analogy to the classical Poisson brackets (\ref{JJ}) and (\ref{JM}) respectively. Now we use \eqref{Jp} to rewrite the Runge-Lenz vector as
\begin{equation}\label{Mp}
\vec{M} = \vec{p}\times\vec{J} -i\vec{p} - \frac{\hat{r}}{2L}\left(L^2H - 2q^2\right).
\end{equation}
The second ambiguity has to do with the position of the factor $\hat{r}$ of the last term. The above choice guarantees that the quantum Runge-Lenz vector commutes with $H$. 

In order to obtain  a Runge-Lenz vector which commutes with the gauged Hamiltonian $H_p$ and still satisfies the relations \eqref{QR},  it turns out that the addition of the term $\vec{f}$ \eqref{f}, which worked in the classical case, also works in the quantum theory. The gauged quantum Runge-Lenz vector is therefore
\bee
\vec{M}^p = \vec{p}\times\vec{J} -i\vec{p} - \frac{\hat{r}}{2L}\left(L^2H_p - 2q^2- pq\right).
\eee
  
A lengthy calculation yields the  commutators 
\begin{align}
\label{commutatorsmj}
[J_i,M^p_j] &= i\epsilon_{ijk}M^p_k,\nonumber \\
[M^p_i,M^p_j] &=  i\left[\frac{1}{L^2}\left(q+\frac{p}{2}\right)^2 - H_p\right]\epsilon_{ijk}J_k,
\end{align}
which quantise the Poisson brackets \eqref{poissonmj}. We also 
find   the following  operator identities:
\begin{align}
\label{identitymj}
\vec{M}^p\cdot\vec{J} &= \vec{J}\cdot\vec{M}^p = -\frac{q}{2L}\left(L^2H_p - 2q^2 - pq\right), \nonumber \\
\vec{M}^p\cdot\vec{M}^p &= \left[H_p - \frac{1}{L^2}\left(q+\frac{p}{2}\right)^2\right](\vec{J}\cdot\vec{J} - q^2 + 1) + \frac{1}{4L^2}\left(L^2H_p - 2q^2- pq\right)^2.
\end{align}

Since the Hamiltonian $H_p$ and the $U(1)$ generator $q$ commute with each other and  all components of  $\vec{M}^p$ and $\vec{J}$, we can fix their eigenvalues and study the commtutation relations of  $\vec{M}^p$ and $\vec{J}$ in a fixed common eigenspace of $H_p$ and $q$. Denoting the eigenvalues by, respectively, $E$ and $-s$, and assuming  the bound state energy range
\begin{equation}\label{tea}
L^2E < \left(s-\frac{p}{2}\right)^2,
\end{equation}
we define the rescaled Runge-Lenz vector,
\begin{equation}\label{rungep}
\tilde{M}^p= \frac{1}{\sqrt{\frac{1}{L^2}(s-\frac{p}{2})^2 - E}}\vec{M}^p.
\end{equation}
Together with the components of $\vec{J}$, it satisfies the  $so(4)$  commutation relations,
\begin{equation}
\label{so4commutators}
[J_i,J_j] = i\epsilon_{ijk}J_k, \ \ [J_i,\tilde{M}^p_j] = i\epsilon_{ijk}\tilde{M}^p_k, \ \ [\tilde{M}^p_i,\tilde{M}^p_j] =  i\epsilon_{ijk}J_k.
\end{equation}

\subsection{Bound states revisited}
The bound state energies of $H_p$ can now be derived from  the isomorphism $so(4)\simeq su(2)\oplus su(2)$ and the  standard representation theory of $su(2)$. We introduce the commuting operators
\begin{equation}
\vec{J}_{\pm} = \frac{1}{2}(\vec{J} \pm \tilde{M}^p),
\end{equation}
and see that  the  two Casimirs 
\begin{equation}\label{c}
J_{\pm}^2 = \frac{1}{4}(\vec{J}\cdot\vec{J} + \tilde{M}^p\cdot\tilde{M}^p) \pm \frac{1}{4}(\tilde{M}^p\cdot\vec{J} + \vec{J}\cdot\tilde{M}^p )
\end{equation}
 have eigenvalues $j_\pm(j_\pm +1)$, where $j_\pm$ are both  non-negative half-integers. Moreover, since $\vec{J}= \vec{J}_+ + \vec{J}_-$, it follows that the total angular momentum quantum number $j$ defined in \eqref{FI} lies in the range
\bee
|j_+ - j_-| \leq j \leq |j_++j_-|. 
\eee
Since $-j\leq s \leq j$, we deduce that 
\bee
\label{srange}
- |j_++j_-| \leq s \leq  |j_++j_-|.
\eee

 In terms of $\tilde{M}^p$, the relations \eqref{identitymj}  read
\begin{align}
\tilde{M}^p\cdot\vec{J} = \vec{J}\cdot\tilde{M}^p &= \frac{s}{2L}\frac{(L^2E -2 s^2 +  ps)}{\sqrt{\frac{1}{L^2}(s-\frac{p}{2})^2 - E}}, \nonumber \\
\tilde{M}^p\cdot\tilde{M}^p +\vec{J}\cdot\vec{J} &=  s^2 - 1 + \frac{(L^2E - 2s^2 + ps)^2}{4((s-\frac{p}{2})^2 - L^2E)}.
\end{align}
Substituting these into \eqref{c} and 
replacing  $
\vec{J}_{\pm}^2 $ by the eigenvalues  $j_\pm(j_\pm +1)$, 
we get two quadratic equations for the unknown\footnote{Note that the definition of $n$ here is consistent with \eqref{lambdaformula} and \eqref{lambdaquant}}
\bee
\label{nE}
n:=\frac{L^2E - 2s^2 + ps}{2L\sqrt{\frac{1}{L^2}(s-\frac{p}{2})^2 - E}},
\eee 
namely
\begin{align}
n^2 + 2s n+ s^2-1 - 4 j_+(j_+ + 1) &= 0, \nonumber \\
 n^2  -2s n+ s^2-1 - 4 j_-(j_- + 1) &= 0.
\end{align}
The roots of the first equation are 
\bee
n = -s \pm (2j_+ + 1),
\eee
and the roots for the second equation are 
\bee
n = s \pm (2j_- + 1). 
\eee
Both equations have to be satisfied for some values of $j_+$ and $j_-$, but combining the upper sign in one with the lower sign in the other implies a value of $s$  which is outside the 
range  \eqref{srange}. Hence, there are only two possible solutions for $n$, one which is manifestly a half-integer  $\geq 1$
\bee
n=
-s + (2j_+ + 1) =s + (2j_- + 1)
\eee
and one which is manifestly a half-integer  $\leq -1$
\bee
n =
-s - (2j_+ + 1) =s - (2j_- + 1).
\eee

Finally solving \eqref{nE}  for $E$ we obtain again the solutions \eqref{SE}  previously obtained via square integrability arguments. However, we still have  four possibilities in total: two choices of sign in \eqref{SE} and two choices for $n$ (positive or negative) 
and in this section  we cannot assume the conditions \eqref{Ebigger} and \eqref{bdcond} to resolve the ambiguity. 
In the four different cases the energy equation \eqref{SE} takes the two possible forms 
\begin{equation}
\label{SEp}
\frac{L^2E}{2} - s^2+ \frac{ps}{2} = 
n^2\left(\pm \sqrt{1+ \frac  { \frac{p^2}{4} -s^2  }   {n^2}}  -1\right).  
\end{equation}
We can eliminate the lower sign because it conflicts with  the lower bound \eqref{quanteniq}. To see this we 
 re-write \eqref{SEp} with the lower sign as 
\bee
L^2E =  -ps -2(n^2-s^2) -  
2n^2\sqrt{1+ \frac  { \frac{p^2}{4} -s^2  }   {n^2}},
\eee
 showing that  $L^2E<  -ps$ in this case. However, this is inconsistent with the energy inequality \eqref{quanteniq} and therefore ruled out.

Turning to the upper sign, we need to consider the two possible signs of $n$ and check the consistency between \eqref{SEp} and \eqref{nE}. In the case  $n\geq 1$, both  \eqref{SEp} and \eqref{nE} assign a positive sign to   $\frac{L^2E}{2} - s^2+ \frac{ps}{2}$ provided $s^2< p^2/4$. In that case  we arrive at the previously derived energy spectrum \eqref{Espectrum} together with the condition \eqref{bdcond} for bound states. However,   $n\leq -1$   is also consistent  provided  $s^2> p^2/4$.  We have not been able to eliminate this case using only the algebraic methods of this section. It seems that the consideration of the actual wavefunction \eqref{eigen} and integrability requirement  \eqref{lambdaquant}  is needed to rule out $n\leq -1$.

Finally turning to the degeneracy of the energy levels, we 
see  that  the  quantum numbers $n$ and $s$ are determined via
\bee
n=j_+ + j_-+1, \quad s= j_+-j_-,
\eee  
and that the degeneracy of the energy level with quantum numbers $n$ and $s$ is the dimension of the tensor product $j_+\otimes j_-$ of the irreducible  angular representations with spins $j_+$ and $j_-$,
\bee
 (2j_+ + 1)(2j_- + 1)=n^2-s^2,
\eee
reproducing and interpreting the degeneracy \eqref{degeneracy} of energy levels.

In this section we have only studied the commutation relations of the angular momentum and Runge-Lenz vectors at energies satisfying  $L^2E< (s-p/2)^2$ and corresponding to bound states. It is not difficult to modify our discussion for the case $L^2E\geq (s-p/2)^2$. The dynamical symmetry algebra of the angular momentum  and of a suitable rescaled Runge-Lenz vector, analogous to \eqref{so4commutators}, turns out to be isomorphic to the Lie algebra $so(3)\ltimes \RR^3$ of the Euclidean group when $L^2E =  (s-p/2)^2$ and isomorphic to the Lie algebra $so(3,1)$  of the Lorentz group when $L^2E >(s-p/2)^2$. In the following section we will see how these three cases can be understood from a unified, geometrical point of view.

\section{Twistorial derivation of the gauged Runge-Lenz vector}

\subsection{Twistors, $SU(2,2)$ symmetry and moment maps}  

It has been known for a while \cite{Moser} that the usual Kepler problem  can be regularised by embedding  momentum three-space into the  three-sphere by means of a stereographic projection.  This  gives a geometrical picture of   the angular momentum and Runge-Lenz vectors as  conserved quantities associated with  symmetries of the round three-sphere. For a full geometrical  understanding of the  dynamical symmetry of the Kepler problem, it is moreover convenient to think of it as the symplectic quotient of an eight dimensional phase space, and the dynamical symmetry algebra for the various energy regimes as subalgebras of $so(4,2)$,  see \cite{GS} for a pedagogical review.

 It was shown in  \cite{CFH}  that one can similarly interpret  angular momentum and Runge-Lenz vectors  of the (ungauged) TN motion as generators of a subalgebra of a $su(2,2)\simeq so(4,2)$ symmetry algebra  acting on an eight-dimensional phase space of twistors. In this section we show how this story can be extended to the gauged  TN dynamics.  We begin with a brief review of the relevant notation. 

For our purposes,  twistor space is $\mathbb{T} = (\mathbb{C}^2\times\mathbb{C}^2)\backslash \{0\}$, and a twistor 
\bee
 Z = \binom{\omega}{\pi} \in \mathbb{T}
\eee
is a  pair of spinors $\omega = \binom{\omega_1}{\omega_2}$ and $\pi = \binom{\pi_1}{\pi_2}$. Twistor space $\mathbb{T}$ is endowed with a pairing
\begin{equation}
(Z,Z) = \bar{\pi}_1\omega_1 + \bar{\pi}_2\omega_2 + \bar{\omega}_1\pi_1 + \bar{\omega}_2\pi_2,
\end{equation}
which can be written as the matrix product $Z^*Z$ where
\begin{equation}
Z^* = Z^{\dagger}A, \ \ A = \left(\begin{array}{cc}
0 & \tau_0 \\
\tau_0 & 0
\end{array}\right),
\end{equation}
is the conjugate spinor and   we write  again  $\tau_0$ for  the $2\times 2$ identity matrix.
The pairing is invariant under $U(2,2)$, but we are particularly interested in the Lie algebra of the subgroup $SU(2,2)$.  Following \cite{CFH}, we pick generators $ \gamma_{KL}$  where the structure constants are purely imaginary, and again write $\tau_i$ for the Pauli matrices:
\begin{align}
\gamma_{0k}& = -\frac{i}{2}\left(\begin{array}{cc}
\tau_k & 0 \\
0 & -\tau_k
\end{array}\right), \ \ 
\gamma_{ij} = \frac{1}{2}\epsilon_{ijk}\left(\begin{array}{cc}
\tau_k & 0 \\
0 & \tau_k
\end{array}\right), \ \ 
\gamma_{06} = \frac{1}{2}\left(\begin{array}{cc}
0 & \tau_0 \\
\tau_0 & 0
\end{array}\right),\nonumber \\
\gamma_{k6} & = \frac{1}{2}\left(\begin{array}{cc}
0 & \tau_k \\
-\tau_k & 0
\end{array}\right), \ \ 
\gamma_{05} = \frac{1}{2}\left(\begin{array}{cc}
0 & \tau_0 \\
-\tau_0 & 0
\end{array}\right), \ \
\gamma_{k5} = \frac{1}{2}\left(\begin{array}{cc}
0 & \tau_k \\
\tau_k & 0
\end{array}\right),
\nonumber \\
\gamma_{56} &= \frac{i}{2}\left(\begin{array}{cc}
\tau_0 & 0 \\
0 & -\tau_0
\end{array}\right), \ \ (\gamma_{LK} = -\gamma_{KL}, \ \ K,L = 0,\dots ,3, 5,6;  \; i,j,k=1,2,3).
\end{align}
For us, two sub-Lie algebras will be important.  The stabiliser Lie algebra of the generator $\gamma_{06}$ is the Lie  algebra generated by 
$s_i:=\frac 1 2 \epsilon_{ijk} \gamma_{jk}$ and $t_i:=\gamma_{5i}$ and  is isomorphic to $so(4)$:
\bee
[s_i,s_j]=i\epsilon_{ijk} s_k, \quad [s_i,t_j] = i\epsilon_{ijk} t_k, \quad  [t_i,t_j] = i\epsilon_{ijk} s_k.
\eee
The stabiliser Lie algebra of the generator $\gamma_{05}$ is the Lie  algebra generated by 
$s_i$ and $r_i:=\gamma_{i6}$ and  is isomorphic to $so(3,1)$:
\bee
[s_i,s_j]=i\epsilon_{ijk} s_k, \quad [s_i,r_j] = i\epsilon_{ijk} r_k, \quad  [r_i,r_j] = -i\epsilon_{ijk} s_k.
\eee

As discussed in \cite{CFH}\footnote{Note that our sign conventions differ from those in  \cite{CFH}},
the space $\mathbb{T}$ has the $U(2,2)$ invariant  one-form
\begin{equation}
\theta =\text{Im}(Z^*dZ) = \frac{1}{2i}( Z^*_{\alpha}dZ^{\alpha}- Z^{\alpha}dZ^*_{\alpha} ),
\end{equation}
whose exterior derivative
\begin{equation}
\label{symform}
\Omega = d\theta = -idZ^*\wedge dZ
\end{equation}
is a symplectic form on $\mathbb{T}$. The $su(2,2)$ generators $\gamma_{KL}$ define vector fields on  $\mathbb{T}$ whose moment maps are 
\begin{equation}\label{conserved}
J_{KL} = Z^{\ast}\gamma_{KL}Z.
\end{equation}

The  diagonal $U(1)$  subgroup  of $U(2,2)$ acts on $\mathbb{T}$,  preserving its symplectic structure. The moment map is $\frac 1 2 Z^* Z$, and the symplectic quotient is the level set 
\begin{equation}
\label{seven}
\mathbb{T}_q=\left\{Z\in \mathbb{T}\left| \frac 1 2 Z^{\ast}Z = q\right.\right\}
\end{equation}
quotiented by the diagonal $U(1)$ action:
 \bee
 \label{twistquot}
 \tilde{\MM}_q = \mathbb{T}_q/ U(1).
 \eee
We now introduce coordinates on $\mathbb{T}$ which are particularly well adapted for describing this quotient.
With the notation  $ \vec{\tau} = (\tau_1,\tau_2,\tau_3)$,  we  parametrise the spinors $\pi$ and $\omega$ in terms of spherical coordinates $(R,\alpha,\beta, \gamma)$ and $\vec{P}\in \RR^3, q\in \RR$ as
\begin{equation}
\label{pi}
\pi = \sqrt{R}\binom{e^{-\frac{i}{2}(\alpha + \gamma)}\cos{\frac{\beta}{2}}}{e^{\frac{i}{2}(\alpha - \gamma)}\sin{\frac{\beta}{2}}},
\end{equation}
and
\bee
\omega =  \left(i\vec{P}\cdot \vec{\tau}+ \frac{q}{R}\tau_0\right)\pi.
\eee

In order to compute  the symplectic structure and moment maps in terms of $\vec{R}$ and $\vec{P}$, we note that
\bee
\pi^\dagger \pi = R, \quad \pi^\dagger \vec{\tau} \pi = \vec{R},
\eee
where 
\bee
\vec{R} = (X_1,X_2,X_3)= (R\sin\beta\cos{\alpha}, R\sin\beta\sin\alpha, R\cos\beta),
\eee
and
\bee
\omega^\dagger \omega = R\vec{P}^2 +\frac{q}{R}, \quad \omega^\dagger \vec{\tau} \omega  =-2\vec{P}\times \vec{J}+\left(R\vec{P}^2 +\frac{q}{R} \right)\hat R.
\eee

It is not difficult to check that, for fixed $q$, the twistor  $Z = \binom{\omega}{\pi} $ satisfies (\ref{seven}) and thus belongs to $\mathbb{T}_q$.  Moreover, the diagonal $U(1)$  acts simply by shifting $\gamma$, so that the vectors  $\vec{P},\vec{R} \in \RR^3$, which are independent of $\gamma$, are good coordinates on the quotient $\tilde{\MM}_q$.

The symplectic structure \eqref{symform} induces a  symplectic structure on $\tilde{\MM}_q$ which can be expressed as  
\begin{align}
\Omega&= dX_l\wedge dP_l +\frac{q}{2R^3}\epsilon_{iln}X_ndX_i\wedge dX_l. 
\end{align} 
The moment maps for $\gamma_{50}$, $\gamma_{60}$ and the generators of their stabiliser Lie algebras can be written in terms of  $\vec{P},\vec{R}$ as 
\begin{align}
Z^*\gamma_{50}Z &= \frac 1 2 ( \omega^\dagger  \omega- \pi^\dagger \pi) =\frac 1 2 \left(
R\vec{P}^2 + \frac{q}{R} -R\right),
 \nonumber \\
Z^*\gamma_{06}Z &= \frac 1 2 (\pi^\dagger \pi +\omega^\dagger  \omega) =\frac 1 2 \left(
R\vec{P}^2 +\frac{q}{R} +R\right),
 \nonumber \\
Z^*\vec{s} Z&=\frac 12 (\omega^\dagger \vec{\tau} \pi + \pi^\dagger \vec{\tau} \omega ) = \vec{R}\times\vec{P} + q\hat{R},  \nonumber \\
Z^*\vec{t} Z&=-\frac 1 2 (\pi^\dagger \vec{\tau} \pi +\omega^\dagger \vec{\tau} \omega)= \vec{P}\times \vec{J}-\frac 1 2 \left(R\vec{P}^2+\frac{q}{R}+R \right)\hat R,
\nonumber \\
Z^*\vec{r} Z&=\frac 1 2 (\pi^\dagger \vec{\tau} \pi -\omega^\dagger \vec{\tau} \omega)
= \vec{P}\times \vec{J}-\frac 1 2 \left(R\vec{P}^2 +\frac{q}{R} -R\right)\hat R.
\end{align}

We can summarise these formulae more neatly by introducing a variable $\kappa$ which can take the values $\pm 1$. Then we write
\begin{equation}
\label{Htilde}
\tilde{H}_p = \frac 1 2 \left(
R\vec{P}^2 + \frac{q}{R} +\kappa R\right),
\end{equation}
for  the moment maps $Z^*\gamma_{50}Z$ and $Z^*\gamma_{06}Z$ with $\kappa=1$ and $\kappa=-1$.
We  write 
\bee
\label{bigJ}
\vec{J} = \vec{R}\times\vec{P} + q\hat R,
\eee
for the vector or moment map $Z^*\vec{s}Z$, and 
finally note that 
\begin{equation}
\label{bigK}
\vec{K} = \vec{P}\times\vec{J} - \tilde{H}_p \hat R
\end{equation}
unifies the moment maps $Z^*\vec{t} Z$ and $Z^*\vec{r} Z$ for $\kappa=1$ and $\kappa=-1$.

It then follows from the general theory of moment maps (and can also be verified directly) that the Poisson brackets of these moment maps are, up to factors of $i$,  the commutators of the Lie algebra elements which  enter the definition. In other words, the brackets are
\begin{align}
\{\tilde{H}_p, J_i\} &= \{\tilde{H}_p, K_i\} = 0 \nonumber \\
\{J_i,J_j\}&=\epsilon_{ijk}J_k,\qquad \{J_i,K_j\}=\epsilon_{ijk}K_k, \qquad \{K_i,K_j\}=\kappa\epsilon_{ijk}J_k.
\end{align}

\subsection{Mapping twistor space to the gauged Taub-NUT phase space }

Adapting the treatment of  \cite{CFH}, we shall show  how the quotient  $\tilde{\MM}_q$ \eqref{twistquot}
can be mapped onto the symplectic quotient $\MM_q$  \eqref{reducedphase} of  the cotangent bundle of TN.  
The map between the phase spaces  is not canonical, but it can be extended to a map on the evolution space,  preserving the presymplectic  (or Poincar\'e-Cartan) two-form. 

For a phase space  $\MM$ with symplectic structure $\omega$ and Hamiltonian $H$, the evolution space is $\MM\times \RR$, and the presymplectic two-form is $\omega +dH\wedge dt$, where $t$ is  a  global (time) coordinate on $\RR$.  The trajectories of the flow with Hamiltonian $H$ can be characterised as the vortex lines of $\omega +dH\wedge dt$, i.e., the lines whose tangent lines are in the null space of  $\omega +dH\wedge dt$. Now consider extended phase spaces $\MM\times \RR$ and $\tilde{\MM} \times \tilde  \RR$ with symplectic structures and Hamiltonians $(\omega,H)$ on $\MM$  and $(\tilde \omega, \tilde H)$  on $ \tilde{\MM}$, and time coordinates $t$ on $\RR$ and $\tilde t$ on $\tilde \RR$.  Then a map 
\bee
F: \MM\times \RR \rightarrow \tilde{\MM} \times  \tilde \RR
\eee
  which satisfies
\bee
F^*(\tilde \omega +d\tilde H\wedge d\tilde t)=\omega +dH\wedge dt
\eee
 will map  trajectories  in the Hamiltonian system  $(\MM,\omega,H)$ to trajectories in the Hamiltonian system  $(\tilde{\MM},\tilde \omega,\tilde H)$. For details and a  pedagogical discussion of Hamiltonian trajectories as vortex lines of Poincar\'e-Cartan structures see \cite{Arnold}.

We should stress that, in contrast to the treatment of the  ungauged case with negative $L$ in \cite{CFH}, and unlike in the usual Kepler problem, no regularisation is required in our case since our Hamiltonian is smooth and finite on the entire phase space.  For definiteness we focus on the case 
\bee
L^2 H_p < \left(q+\frac{p}{2}\right)^2,
\eee
 which is relevant for bounded orbits.

The two extended phase spaces we would like to map into each other are $\MM_q\times \RR$  with presymplectic two-form 
\begin{equation}
\sigma =\omega + dH_p\wedge dt =  dx_l\wedge dp_l + \frac{q}{2r^3}\epsilon_{iln}x_ndx_i\wedge dx_l + dH_p\wedge dt,
\end{equation}
and  $ \tilde{\MM}_q\times\tilde \RR$
with presymplectic two-form 
\begin{equation}
\Sigma = dX_l\wedge dP_l + \frac{q}{2R^3}\epsilon_{iln}X_ndX_i\wedge dX_l + d\tilde{H}_p\wedge d\tilde t.
\end{equation}
The required  map is most easily written down in terms of the coordinates $(\vec{P},\vec{R},\tilde t)$ of $ \tilde{\MM}_q\times \tilde \RR$ and the coordinates $(\vec{p},\vec{r},t)$ on  $\MM_q\times \RR$. It takes the form  
\begin{align}
 F: \MM_q\times \RR  \rightarrow  \tilde{\MM}_q \times \tilde \RR,  \qquad  (\vec{r},\vec{p},t) \mapsto (\vec{R},\vec{P},\tilde t),
 \end{align}
 where
\begin{align}
\label{newc}
   \vec{R} & = \vec{r}\sqrt{\frac{1}{L^2}\left(q+\frac{p}{2}\right)^2 - H_p}, \nonumber \\
\vec{P} & = \frac{\vec{p}}{\sqrt{\frac{1}{L^2}\left(q+\frac{p}{2}\right)^2 - H_p}}, \nonumber \\
\tilde t & = \frac{\sqrt{\frac{1}{L^2}\left(q+\frac{p}{2}\right)^2 -H_p}}{\frac{p^2}{L} + \frac{pq}{2L} - \frac{1}{2}LH_p}\left\{\vec{r}\cdot\vec{p} + 2\left[\frac{1}{L^2}\left(q+\frac{p}{2}\right)^2 -H_p\right]t\right\}.
 \end{align} 
 
A lengthy calculation shows 
\begin{align}
\label{pull1}
F^*(dX_l\wedge dP_l) & = dx_l\wedge dp_l + \frac{\frac{1}{2}(\vec{p}\cdot d\vec{r} + \vec{r}\cdot d\vec{p})\wedge dH_p}{\frac{1}{L^2}(q+\frac{p}{2})^2 -H_p}, \nonumber \\
F^*\left(\frac{q}{2R^3}\epsilon_{iln}X_ndX_i\wedge dX_l\right)& = \frac{q}{2r^3}\epsilon_{iln}x_ndx_i\wedge dx_l,
\end{align}
and 
\begin{equation}
F^*(d\tilde{H}_p\wedge d\tilde t) = dH_p\wedge dt -  \frac{\frac{1}{2}(\vec{p}\cdot d\vec{r} + \vec{r}\cdot d\vec{p})\wedge dH_p}{\frac{1}{L^2}\left(q+\frac{p}{2}\right)^2 -H_p}.
\end{equation}
Combining these, 
 we deduce
\begin{equation}
F^*\Sigma = \sigma,
\end{equation}
as claimed. It follows that $F$ maps solutions of the Hamilton equations
\bee
\frac{d\vec{r}}{dt} =\frac{\partial H_p}{\partial  \vec{p}}, \qquad  \frac{d\vec{p}}{dt} = -\frac{\partial H_p}{\partial \vec{r}}
\eee
to solutions of the Hamilton equations
\bee
\frac{d\vec{R}}{d\tilde t} = \frac{\partial \tilde  H_p}{\partial  \vec{P}}, \qquad  \frac{d\vec{P}}{d\tilde t} = -\frac{\partial \tilde H_p}{\partial \vec{R}}.
\eee

Having seen that $F$ maps trajectories to trajectories, albeit traversed at different rates, we conclude this section by showing how $F$ relates observables. 
It is easy to check that  $F$ maps the angular momentum in $\tilde{\MM}_q$ to the  the angular momentum in $\MM_q$, i.e., the substitution of \eqref{newc} into \eqref{bigJ} gives the TN angular momentum 
\begin{equation}
\vec{J} = \vec{r}\times\vec{p} + q\hat{r}.
\end{equation} 
The Hamiltonians and the Runge-Lenz generators, however, are related by pulling back with $F$ together with rescaling.
Substituting the expressions  \eqref{newc} into  the Hamiltonian \eqref{Htilde}   with $\kappa =1$,
one finds the re-scaled  gauged TN Hamiltonian,
\begin{equation}\label{Ham}
\tilde{H}_p = \frac{ L^2 H_p - 2q^2- pq  }{2L\sqrt{\frac{1}{L^2}(q+\frac{p}{2})^2 - H_p}},
\end{equation}
and the substitution into \eqref{bigK} (again with $\kappa=1$) gives a rescaled Runge-Lenz vector \eqref{rungep},
\begin{equation}
\vec{K} = \frac{\vec{p}\times\vec{J} - \frac{1}{2L} (L^2H_p - 2q^2 - pq)\hat{r}}{\sqrt{\frac{1}{L^2}(q+\frac{p}{2})^2 - H_p}}.
\end{equation}
     
\section{Discussion and Conclusion}

Our results show that the inclusion of the  magnetic field \eqref{A}  in the discussion of dynamics on TN space  is both  mathematically natural and phenomenologically  interesting. The gauge field   preserves  the  geometrical $U(2)$ symmetry of  TN and here we showed that it also preserves the dynamical symmetry of the associated phase space.  

As suggested by our qualitative discussion of magnetic binding in Sect.~2, the form of the magnetic field ensures the existence of bounded orbits classically    and  of bound states  quantum mechanically. As a result, TN dynamics with the  magnetic field \eqref{A} combines aspects of two  paradigmatic  systems  of classical and quantum mechanics - the Kepler problem and the Landau problem of a charged particle in a magnetic field - into one model that preserves and merges the most interesting features of both.

In fact, one can  understand  many results of this paper qualitatively by  thinking of  gauged TN dynamics  as a combination of a  Landau problem in the  cigar-shaped submanifolds of  TN space, reviewed in Sect.~\ref{TNreview},  and  a Kepler problem on the base. The magnetic field on the cigars acts as `magnetic plug' which keeps  trajectories in a bounded region  and produces quantum bound states, provided the energy is sufficiently small and  $|q|< |p/2|$.  This  last condition is precisely the condition for bounded trajectories in our toy model of Sect.~2.
A further link with Landau states,  explained in \cite{AFS},  is that in  the limit  where the TN parameter $\epsilon$ in \eqref{solu} (which we set to 1 in our discussion) is taken to zero, the gauged TN problem actually becomes  a four-dimensional Landau problem. 

Some of the details of our results deserve further comments.  
The spectrum of quantum bound states \eqref{Espectrum} depends both on the magnetic flux parameter  $p$ and the   quantum number $s$.  Importantly, it also depends on the relative sign of those two quantum numbers through the appearance of the product $ps$ in  the energy levels \eqref{Espectrum}. This is a consequence of the  breaking of the discrete symmetry \eqref{broken}. In view of the interpretation of $s$ as a component of the body-fixed angular momentum, the dependence on energy levels on the alignment of this angular momentum with the magnetic flux may is essentially an (anomalous) Zeeman effect.

In the interpretation of the TN manifold as a collective coordinate space of magnetic monopoles in \cite{GM, Gauntlett, Weinberg}, 
the body-fixed angular momentum quantum number $s$ is an electric charge. Adopting that nomenclature, and recalling that the parameter $p$ measure the magnetic flux of an external magnetic field, the equality of the purely magnetic scattering cross section \label{magscat} and the purely electric cross section  \label{elecscat}  may be viewed as a manifestation of a magnetic-electric duality. As far as we are aware, this is not expected in this context, and it remains an open problem to find a simple explanation of it. 

\noindent {\bf Acknowledgments} \;\; RJ thanks MACS at  Heriot-Watt University for a PhD scholarship. BJS acknowledges support through the EPSRC grant `Dynamics in Geometric Models of Matter'. We  thank the Isaac Newton Institute for hospitality during the final stage  of writing this paper  and Guido Franchetti for producing the plots in Fig.~1.

\end{document}